\documentclass[%
 reprint,
superscriptaddress, 
showpacs,preprintnumbers,
 amsmath,amssymb,
 aps,
prstab,
]{revtex4-1}

\usepackage{graphicx}
\usepackage{dcolumn}
\usepackage{bm}

\newcommand{\degree}{$^{\circ}$}
\newcommand{\qz}{$Q_0$}
\begin{document}


\title{Effect of high temperature heat treatments on the quality factor of a large-grain superconducting radio-frequency niobium cavity}

\author{P. Dhakal}
\affiliation{Thomas Jefferson National Accelerator Facility, Newport News, Virginia 23606, USA}
\author{G. Ciovati}
\affiliation{Thomas Jefferson National Accelerator Facility, Newport News, Virginia 23606, USA}
\author{G. R. Myneni}
 \email{Corresponding author.\\rao@jlab.org}
\affiliation{Thomas Jefferson National Accelerator Facility, Newport News, Virginia 23606, USA}
\author{K. E. Gray}
\affiliation{Material Science Division, Argonne National Laboratory, Argonne, Illinois 60493, USA}
\author{N. Groll}
\affiliation{Material Science Division, Argonne National Laboratory, Argonne, Illinois 60493, USA}
\author{P. Maheshwari}
\affiliation{North Carolina State University, Raleigh, North Caolina 27695, USA}
\author{D. M. McRae}
\affiliation{National High Magnetic Field Laboratory, Florida State University, Tallahassee, Florida 32310, USA}
\author{R. Pike}
\affiliation{College of William and Mary, Williamsburg, Virginia 23187, USA}
\author{T. Proslier}
\affiliation{Material Science Division, Argonne National Laboratory, Argonne, Illinois 60493, USA}
\author{F. Stevie}
\affiliation{North Carolina State University, Raleigh, North Caolina 27695, USA}
\author{R. P. Walsh}
\affiliation{National High Magnetic Field Laboratory, Florida State University, Tallahassee, Florida 32310, USA}
\author{Q. Yang}
\affiliation{Center for Materials Research, Norfolk State University, Norfolk, Virginia 23504, USA}
\author{J. Zasadzinzki}
\affiliation{Physics Department, Illinois Institute of Technology, Chicago, Illinois 60616, USA}

\date{\today}

\begin{abstract}
Large-grain Nb has become a viable alternative to fine-grain Nb for the fabrication of superconducting radio-frequency cavities. In this contribution we report the results from a heat treatment study of a large-grain 1.5 GHz single-cell cavity made of "medium purity" Nb. The baseline surface preparation prior to heat treatment consisted of  standard buffered chemical polishing. The heat treatment in the range 800 - 1400~\degree C was done in a newly designed vacuum induction furnace. $Q_0$ values of the order of $2\times10^{10}$ at 2.0~K and peak surface magnetic field ($B_p$) of 90~mT were achieved reproducibly. A $Q_0$-value of $(5 \pm 1)\times10^{10}$ at 2.0~K and $B_p$ = 90~mT was obtained after heat treatment at 1400~\degree C. This is the highest value ever reported at this temperature, frequency and field. Samples heat treated with the cavity at 1400~\degree C were analyzed by secondary ion mass spectrometry, secondary electron microscopy, energy dispersive X-ray, point contact tunneling and X-ray diffraction and revealed a complex surface composition which includes titanium oxide, increased carbon and nitrogen content but reduced hydrogen concentration compared to a non heat-treated sample.   
\end{abstract}

\maketitle


\section{\label{sec1}Introduction}
Since its introduction in 2005 as a material potentially suitable for superconducting radio-frequency (SRF) cavities for accelerators, multi-cell cavities made of large-grain Nb obtained directly from an ingot have demonstrated comparable performance, both in terms of accelerating gradient, $E_{acc}$, and $Q_0$-values, to those made of standard high purity (residual resistivity ratio, $RRR>300$), fine-grain (ASTM $>$ 5) niobium, for the same "standard" surface treatment procedures. Two large-grain nine-cell 1.3 GHz cavities have been operating in the FLASH accelerator at DESY with $E_{acc}$-values of $\sim$30~MV/m \cite{Singer2010}, while eight new additional cavities, which will be installed in a cryomodule for the XFEL accelerator also at DESY, achieved $E_{acc}$-values of up to 45.6~MV/m in a vertical test at 2.0~K \cite{Reschke2011}.

The experience with large-grain cavities at several laboratories showed that 10-30\% higher \qz\ values at medium field ($E_{acc} \sim$20~MV/m) could be obtained compared to fine-grain cavities, for the same frequency and temperature \cite{Ciov2010,Geng2011,Singer2013}. While the causes for the observed higher Q-value are not known, some possibilities include fewer grain boundary losses, higher thermal conductivity at 2.0~K due to the so-called "phonon-peak", lower flux-trapping efficiency and lower hydrogen pick-up.

In the 1970s, heat treatment in ultra-high vacuum (UHV) at $\sim$1800~\degree C for $\sim$10~h was the standard practice to achieve high \qz\ values in bulk Nb cavities \cite{Pfister1976}. The main benefits were associated to the dissolution of precipitates and clusters of impurities and to "stress relief", therefore greatly reducing the dislocation content. In the 1990s, the heat treatment temperature was reduced to $\sim$1300~\degree C and a solid state getter, such as Titanium, was used inside the furnace to "post-purify" the Nb cavities \cite{Kneisel1988}. This was done mainly to increase the thermal conductivity and therefore provide better thermal stabilization of possible defects in the Nb inner surface. In the last decade, the heat treatment temperature has been reduced to $\sim$600-800~\degree C, mainly just to degas hydrogen absorbed by the Nb during cavity fabrication and surface treatments \cite{Knobl2002}. 

Removal of $\sim$20~$\mu$m of material from the inner cavity surface by either Buffered Chemical Polishing (BCP) or Electropolishing (EP) is a standard step following the heat treatment and allows eliminating a "polluted" layer which contains high concentration of impurities absorbed into Nb from the residual gases in the furnace, during cool-down to room temperature. Nevertheless, the chemical processing readily introduces hydrogen, which segregates at the Nb surface and  has been known to degrade the rf superconducting properties of Nb cavities \cite{Bonin1991, Isagawa1980}.

Increasing the quality factor of SRF cavities is one of the key factors to enable efficient continuous wave (CW) linear accelerators (linacs) of various kind (energy recovery linacs, free electron lasers, high-intensity proton linacs) for many different applications, ranging from light sources to energy production via accelerator-driven reactors. A "streamlined" treatment procedure to be applied to large-grain cavities has been proposed to obtain improved \qz-values at a reduced cost \cite{Ciov2010}. The bulk material removal after fabrication should be done by Centrifugal Barrel Polishing (CBP) to smooth the equatorial weld area with an "environmentally friendly" process, followed by a smaller amount of material removal by BCP. A high-temperature heat-treatment in a clean UHV furnace should then be applied, followed by standard high-pressure rinse (HPR) with ultra-pure water to remove particulates from the surface, before assembly in an ISO 4 clean room.
In order to determine the optimum heat treatment temperature to maximise \qz(2.0~K, 90~mT), the \qz($B_p$) of a large-grain single-cell cavity was measured at 2.0 K after heat treatment at different temperatures, in the range 800-1400~\degree C for 3-6~h, in a newly built induction furnace with all-niobium hot zone. Approximately 20~$\mu$m were etched by BCP 1:1:2 after each heat treatment to provide a new "baseline" surface. Additional characterization techniques such as optical inspection of the cavity inner surface and thermal mapping were applied for each test, besides the rf measurements of \qz($T$) and \qz($B_p$). Samples heat treated with the cavity have been analyzed by Secondary Ion Mass Spectrometry (SIMS), Secondary Electron Microscopy (SEM), Energy Dispersive X-ray (EDX) and Point Contact Tunneling (PCT) to determine the impurity concentration and depth profiles as well as the electronic density of states in the superconducting state.

\section{\label{sec2}Cavity preparation and treatment}
The single-cell cavity was fabricated from 3.125~mm thick discs sliced by wire electro-discharge machining from an ingot (labelled "ingot G") supplied by CBMM, Brazil. The Tantalum content of the ingot is $\sim$1375~wt.ppm and the RRR, obtained from the thermal conductivity at 4.2 K of a sample from the same ingot, is $\sim$200. The cavity was built with the standard fabrication method, consisting of deep-drawing of half-cells joined at the equator and to the cut-off tubes by electron beam welding. The cell shape is that of the center-cell of the cavity installed in the CEBAF accelerator. The resonant frequency of the TM$_{010}$ mode in the single-cell is 1.47~GHz, the ratios $B_p/E_{acc}$ and $E_p/E_{acc}$ are 4.43~mT/(MV/m) and 1.78, respectively, and the geometry factor, \textit{G}, is 273~$\Omega$ .

\begin{figure}
\includegraphics{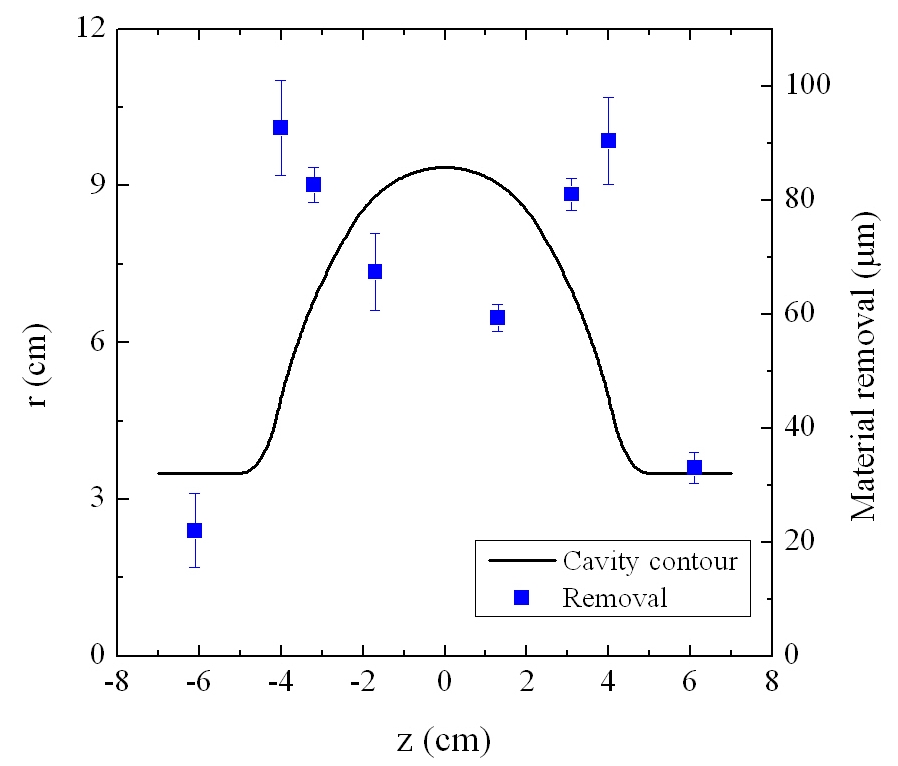}
\caption{\label{fig:removal} Material removal by CBP measured at different locations along the cavity contour \textit{r}(\textit{z}).}
\end{figure}

An average of 73 $\pm$ 13~$\mu$m was removed from the cavity inner surface by CBP after fabrication. This was done in three steps with different media: a "coarse" 5.5~h long polishing step with a cermic mix media, a "medium" 10~h long polishing step with plastic cones media and a "fine" 12~h long polishing step with corn cobs. The rotation speed for each step is 115~rpm. Figure ~\ref{fig:removal}
shows a plot of the total material removal at different locations along the cavity profile. Thickness measurements were done with an ultrasonic probe.

After CBP, an average of 65 $\pm$ 9~$\mu$m were removed from the inner cavity surface by BCP 1:1:2. The cavity was degreased in a solution of ultra-pure water and detergent, with ultrasonic agitation, then it was rinsed, dried and loaded in the induction furnace. The furnace features an all Nb hot-zone consisting of a can and a vertical stand which holds the cavity inside the can's volume. The outside of the Nb can is contained in a quartz tube and an induction coil sits on the outside of the quartz tube. The volumes between the quartz tube and the can and inside the can are separated and part of different vacuum systems. The Nb can is heated by induction, whereas the cavity is heated by irradiation from the hot Nb can. Further details about the induction furnace can be found in \cite{Dhakal2012}.

The cavity was heat treated at 800~\degree C for 3~h, then cooled to $\sim$160~\degree C, held for $\sim$7~h, and finally cooled down to room temperature. The furnace was vented with ultra-pure nitrogen gas at $\sim$150 \degree C. The average total pressure at 800~\degree C was $3\times10^{-5}$~Torr. After heat treatment (HT), the cavity was degreased, HPRed for 1~h, dried for $\sim$3~h in an ISO 4 clean room. Stainless steel blanks with pump-out port and RF antennae were assembled on the cavity flanges with indium wire as gasket. The cavity was then evacuated to $\sim10^{-8}$~mbar on a vertical test stand which is inserted in a cryostat for rf testing in a liquid He bath.
The subsequent treatment procedures were as follows: (i) $\sim$20~$\mu$m material removal by BCP 1:1:2 (baseline 1); (ii) HT at 800~\degree C for 6~h at an average total pressure, $P_{avg}$, of $7.6\times10^{-6}$~Torr; (iii) low-temperature baking (LTB) at 120~\degree C for 24~h; (iv) $\sim$15~$\mu$m material removal by BCP 1:1:2 (baseline 2); (v) HT at 1000~\degree C for 6~h at $P_{avg} \cong 3.3\times10^{-6}$~Torr; (vi) LTB at 120~\degree C for 12~h; (vii) $\sim$29~$\mu$m material removal by BCP 1:1:2 (baseline 3); (viii) HT at 1200~\degree C for 6~h at $P_{avg} \cong 2.5\times10^{-6}$~Torr; (ix) LTB at 120~\degree C for 12~h; (x) $\sim$20~$\mu$m material removal by BCP 1:1:2 (baseline 4); (xi) HT at 1400~\degree C for 3~h at $P_{avg} \cong 9.6\times10^{-6}$~Torr; (xii) LTB at 120~\degree C for 12~h. The cavity treatments are summarized in Table~\ref{tab:tabl1}. 

\begin{table*}
\caption{\label{tab:tabl1} Summary of cavity treatments including the material removal by BCP, the temperature/duration and average pressure during HT, the furnace venting gas, the temperature/duration of the LTB and the figure showing the \qz vs. $B_p$ results.
}
\begin{ruledtabular}
\begin{tabular}{cccccc}
 Removal ($\mu$m)&HT&$P_{avg}$ (Torr)&Venting&LTB&Fig.\\ \hline
 65&800 \degree C$/$3 h&$3\times10^{-5}$&N$_2$&160 \degree C$/$7 h&4 \\
 20&800 \degree C$/$6 h&$7.6\times10^{-6}$&O$_2$&120 \degree C$/$24 h&4 \\
 15&1000 \degree C$/$6 h&$3.3\times10^{-6}$&O$_2$&120 \degree C$/$12 h&5 \\
 29&1200 \degree C$/$6 h&$2.5\times10^{-6}$&O$_2$&120 \degree C$/$12 h&6 \\
 20&1400 \degree C$/$3 h&$9.6\times10^{-6}$&O$_2$&120 \degree C$/$12 h&7 \\
 \end{tabular}
\end{ruledtabular}
\end{table*}

\begin{figure}
\includegraphics{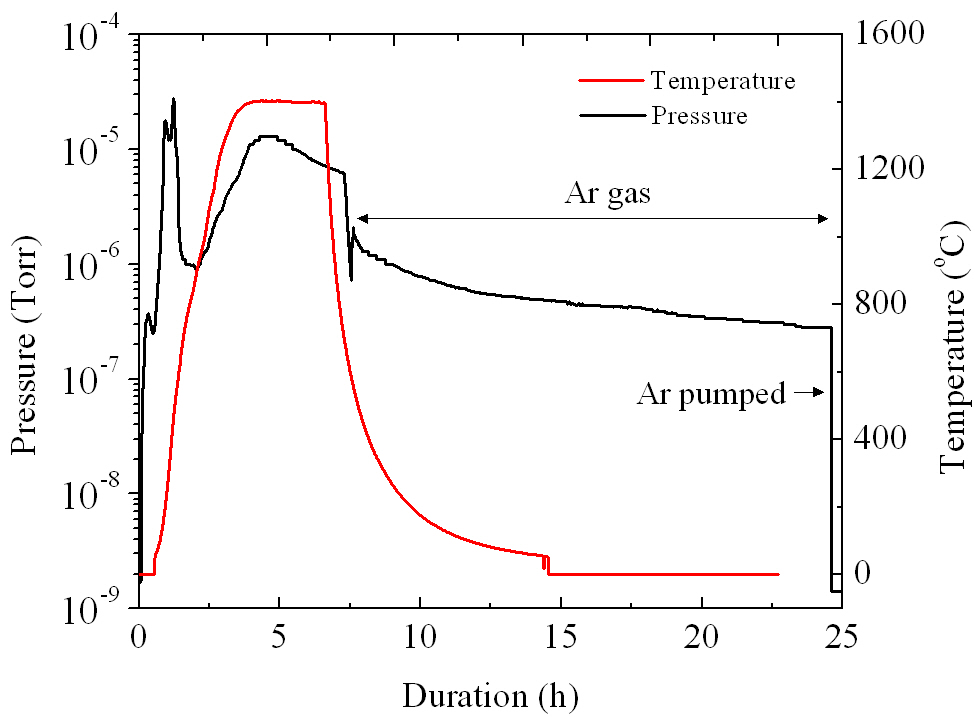}
\caption{\label{fig:heatt} Temperature and total pressure during heat treatment at 1400~\degree C. During cooldown the total pressure was kept at $\sim10^{-6}$~Torr by admitting ultra-high purity Argon gas. The temperature was measured with a type C thermocouple using a voltmeter with $\sim$1~mV resolution ($\sim$70~\degree C).}
\end{figure}

\begin{figure}
\includegraphics{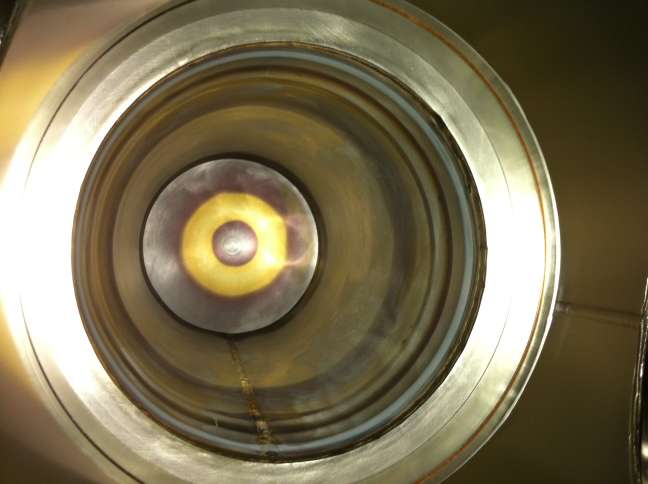}
\caption{\label{fig:ring} Picture of the inside of the Nb susceptor can of the induction furnace showing the presence of a "gold colored" ring at the top after the HT at 1400~\degree C.}
\end{figure}

The residual gases in the furnace during heat treatment were dominated by hydrogen. During the HT cycles in steps (ii), (v), (viii), (xi) the total pressure during the furnace cool-down was maintained at $\sim10^{-6}$~Torr by flowing ultra-high purity Argon gas into the chamber through a leak-valve. This was done to mitigate pollution of the Nb surface during cool-down \cite{Faber1972}. Once room temperature was reached, Argon was pumped out and the furnace was vented to 1 atm, with ultra-high purity Oxygen gas to grow a so-called "dry oxide" layer. For all heat treatments, the heating rate was $\sim$5~\degree C/min and the cooling time was $\sim$8~h. Figure ~\ref{fig:heatt} shows the total pressure and temperature during HT at 1400~\degree C.
The total pressure in the furnace before heat treatments is $\sim 2\times10^{-9}$~Torr, except prior to the HT at 1400~\degree C, when the pressure was $\sim 2\times10^{-8}$~Torr. In this case, residual gases included nitrogen (partial pressure $\sim 1\times10^{-9}$~Torr), oxygen (partial pressure $\sim 5\times10^{-10}$~Torr) and a specie with an atomic mass unit of 69 (partial pressure $\sim 1\times10^{-9}$~Torr). During the heat treatment at 1400~\degree C, the flange of the Vacuum Coupling Radiation fitting for the thermocouple inserted in the furnace was tightened few times and the residual gases in the furnace after the HT were only hydrogen and water vapour. It was noticed that, after the HT at 1400~\degree C, a "gold" colored ring was formed on the top of the Nb susceptor can of the induction furnace, as shown in Fig.~\ref{fig:ring}.

The LTB was done by enclosing the cavity, under vacuum and connected to the vertical test stand, in an insulating box and flowing hot Nitrogen gas into the box. The heating rate is $\sim$0.3~\degree C/min, the cooling time is $\sim$4~h. $P_{avg}$ at 120~\degree C was $\sim1\times10^{-7}$~Torr, mostly Hydrogen.

\section{\label{sec3}Cavity measurement results}
The rf tests consisted of measurements of \qz(\textit{T}, $\sim$10~mT) during the pump-down of the helium bath from 830~Torr (4.3~K) to as low as 4~Torr (1.5~K) and of \qz(2.0~K, $B_p$). These measurements were done after each treatment step listed in the previous section. The input and pick-up antennae have fixed coupling to the cavity with a $Q_{ext}$-value of $\sim1.2\times10^{10}$ and $\sim7\times10^{11}$, respectively. A thermometry system \cite{Ciov2005}, consisting of 576 carbon resistance-temperature-devices in contact with the outer cavity surface, was routinely used during rf testing at 2.0 K to obtain temperature maps of the outer cavity surface as a function of $B_p$. In the following, we identify a thermometer position by a pair (angle, sensor No.), where the angle is the azymuthal location of a thermometry board and sensors No. 1 to 16 follow the cavity contour, along the main axis. Sensors No. 1 and No. 16 are located at the top and bottom cut-off tubes, close to the irises, respectively, whereas sensors No. 8 and No. 9 are located just above and below the equator weld, respectively.
Following the rf test, an optical inspection of the inner surface at quench locations was done using a high-resolution camera \cite{Iwashita2008}.

\begin{figure}
\includegraphics{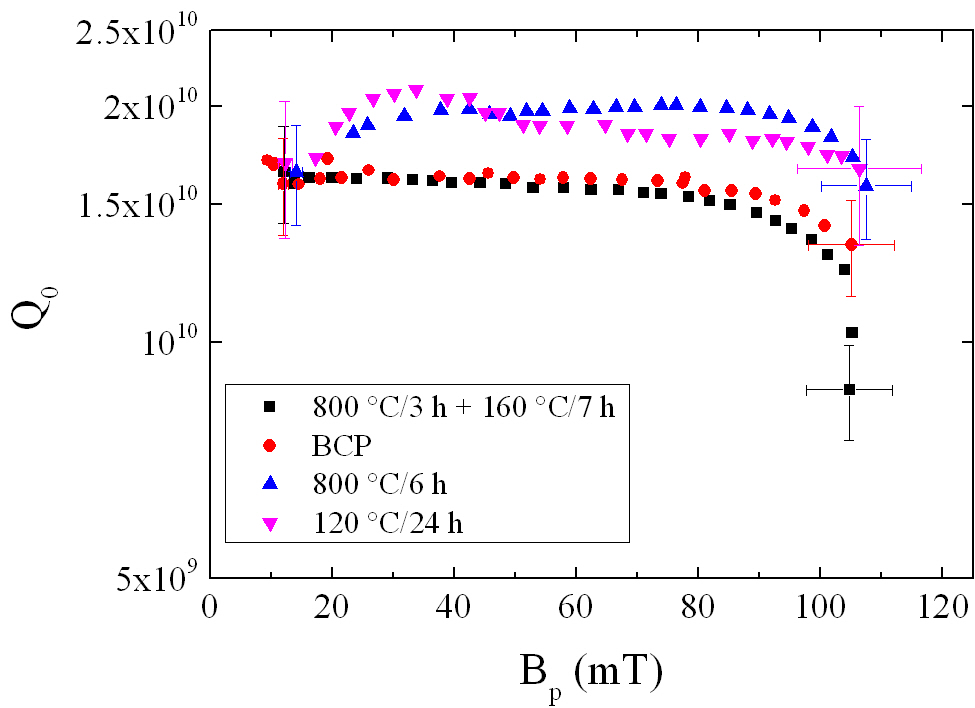}
\caption{\label{fig:q1} \qz(2.0 K) vs. $B_p$ measured after the first HT at 800~\degree C, followed by BCP, by the second HT at 800~\degree C and by LTB.}
\end{figure}

Figure~\ref{fig:q1}
shows a plot of \qz\ vs. $B_p$ measured at 2.0~K after the first HT at 800~\degree C, after BCP, after the second HT at 800~\degree C and after LTB. The cavity was limited by a strong multipacting at $\sim$105~mT after the first test, by a quench at $\sim$105~mT at location (60, 9-10) after the following three tests. \qz(2.0~K, 90~mT) increased from (1.6 $\pm$ 0.2)$\times 10^{10}$ after BCP up to (2.0 $\pm$ 0.3)$\times 10^{10}$ after HT at 800~\degree C for 6~h. LTB reduced the Bardeen-Cooper-Schrieffer (BCS) surface resistance, $R_{BCS}$, at 4.3~K by $\sim38\%$ but doubled the residual resistance, $R_{res}$, from $\sim$3 to $\sim$6~n$\Omega$.

\begin{figure}
\includegraphics{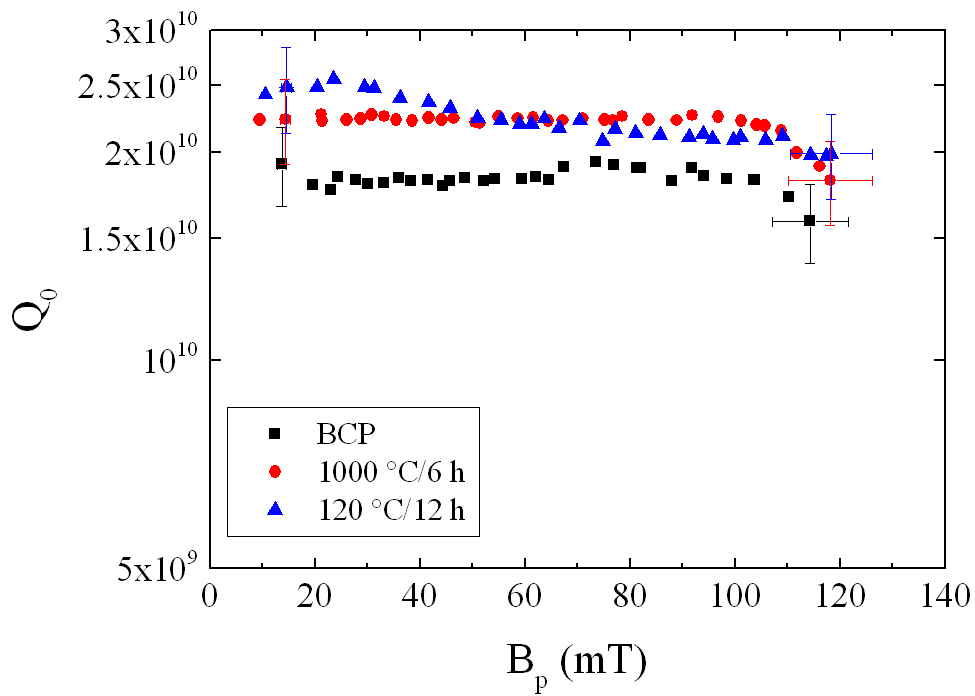}
\caption{\label{fig:q2} \qz(2.0K) vs. $B_p$ measured after etching by BCP, followed by HT at 1000~\degree C and by LTB.}
\end{figure}

Figure~\ref{fig:q2}
shows a plot of \qz\ vs. $B_p$ measured at 2.0~K after additional BCP, after HT at 1000~\degree C and after LTB. In this series of tests the cavity was limited by quench at $\sim$114-118~mT occurring at location (310, 9). \qz(2.0~K, 90~mT) increased from (1.9 $\pm$ 0.3)$\times 10^{10}$ after BCP up to (2.2 $\pm$ 0.3)$\times 10^{10}$ after HT at 1000~\degree C. LTB for shorter time (12~h instead of 24~h) limited the increase in $R_{res}$ to $\sim$0.6~n$\Omega$.

\begin{figure}
\includegraphics{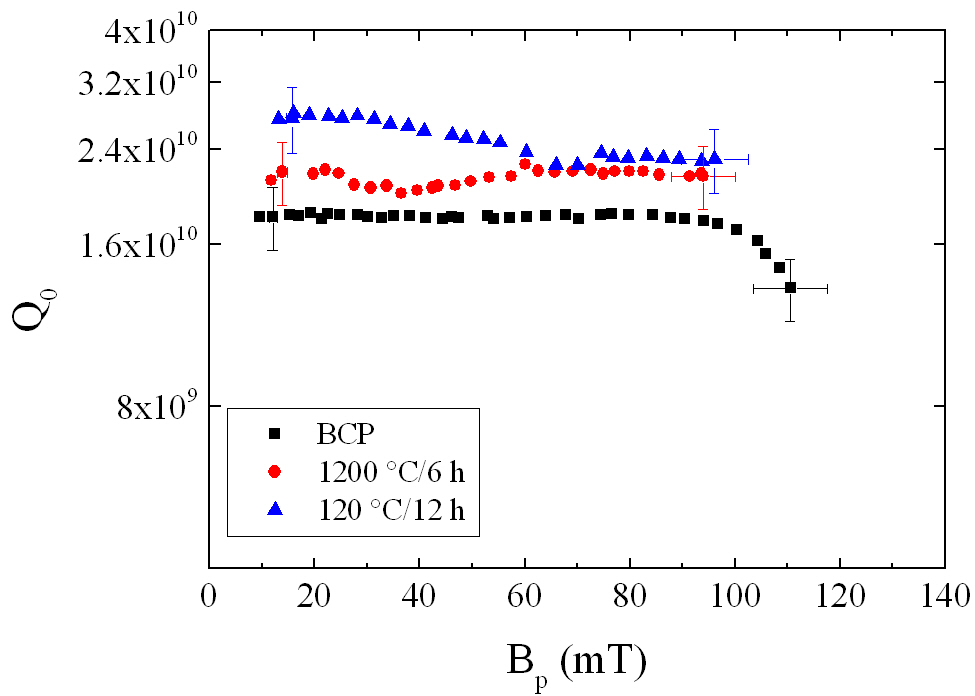}
\caption{\label{fig:q3} \qz(2.0 K) vs. $B_p$ measured after etching by BCP, followed by HT at 1200~\degree C and by LTB.}
\end{figure}

Figure~\ref{fig:q3}
shows a plot of \qz\ vs. $B_p$ after additional BCP, after HT at 1200~\degree C and after LTB. The cavity was limited by quench at $\sim$110~mT after BCP and by multipacting induced quench at $\sim$96~mT at location (220, 11) after HT and LTB. Multipacting at $\sim$90~mT was also observed during the test after BCP. \qz(2.0~K, 90~mT) increased from (1.8 $\pm$ 0.2)$\times 10^{10}$ after BCP up to (2.3 $\pm$ 0.3)$\times 10^{10}$ after LTB. The residual resistance after HT was less than 1~n$\Omega$ and the \qz (1.5~K, 20~mT) was greater than $1\times 10^{11}$.

\begin{figure}
\includegraphics{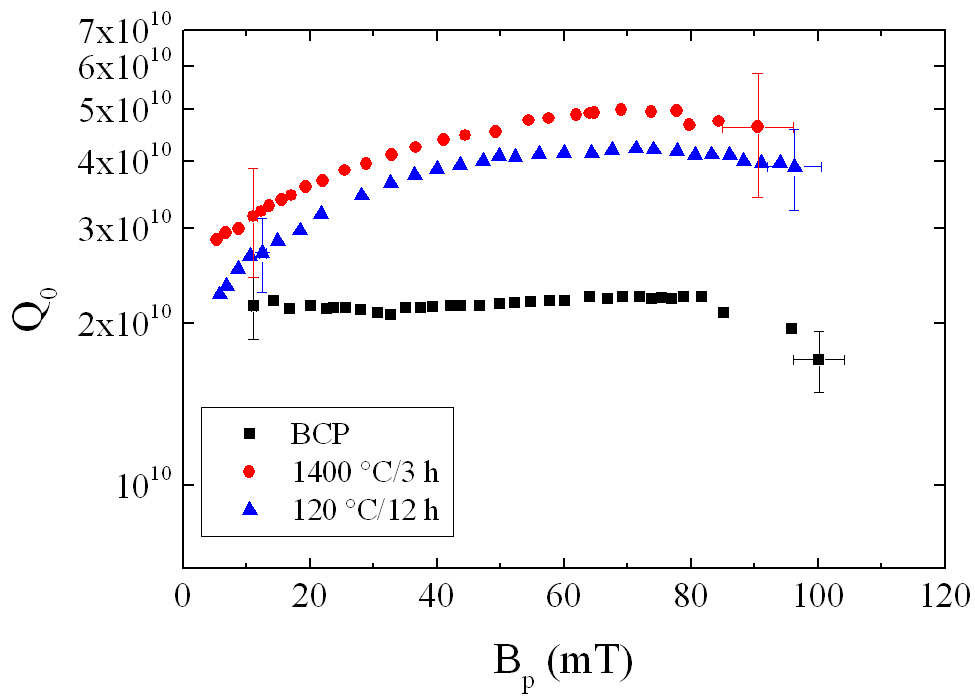}
\caption{\label{fig:q4} \qz(2.0 K) vs. $B_p$ measured after etching by BCP, followed by HT at 1400~\degree C and by LTB. A \qz-value of $(5~\pm~1)~\times~10^{10}$ was obtained at 90~mT.}
\end{figure}

\begin{figure}
\includegraphics{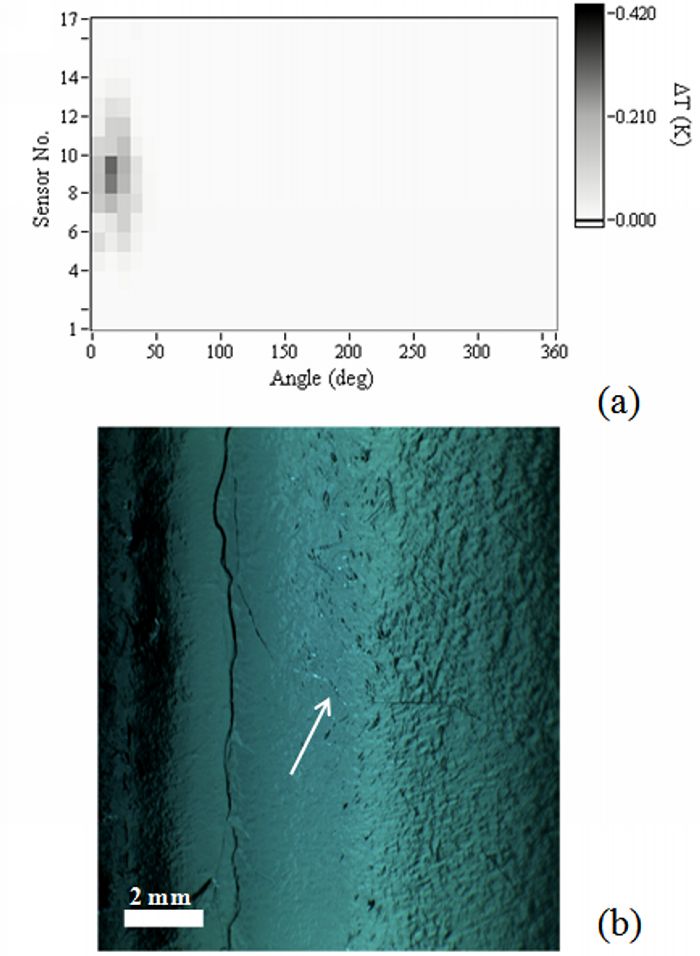}
\caption{\label{fig:quench} "Unfolded" temperature map showing the temperature increase, $\Delta T$, above the He bath at 2.0~K during quench at 91~mT after heat treatment at 1400~\degree C (a) and image of the inner cavity surface at the quench location showing the presence of a grain boundary (b).}
\end{figure}

Figure~\ref{fig:q4}
shows a plot of \qz\ vs. $B_p$ after additional BCP, after HT at 1400~\degree C and after LTB. The cavity was limited by quench at $\sim$100~mT at location (30, 9) after BCP and by quench at location (10, 8) after HT. Multipacting was observed during the test after BCP at $\sim$86~mT. $R_{BCS}$ at 4.3~K was reduced by $\sim24\%$ after HT at 1400~\degree C. A $R_{res}$ of $\sim$1~n$\Omega$ was also obtained after HT which, combined with the reduced $R_{BCS}$, resulted in a low-field \qz-value of $\sim2\times 10^{11}$ at 1.5~K. The \qz\ vs. $B_p$ dependence at 2.0~K shows an increase of \qz\ up to $\sim$60~mT, resulting in a \qz-value of (4.6 $\pm$ 1.0)$\times 10^{10}$ at 90~mT. To our knowledge this is the highest \qz-value ever reported at this field, frequency and temperature. The temperature maps as a function of field taken at 2.0~K after HT showed no significant heating up to the quench field, as the temperature difference from the He bath, $\Delta$\textit{T}, remained below 1 mK. The rf test after HT was done twice, without warming up the cavity, with two different RF systems and by two different operators and the data were within 5\%. It was also verified that there was no significant dependence of the insertion loss of the input power cable on rf power, up to 1.5~W, which was the incident power needed to quench the cavity, as that could introduce a \qz-dependence on field.
The cavity optical inspections did not show any outstanding feature at quench locations, except after 1400~\degree C heat treatment, when the quench location was found to be at a grain boundary, as shown in Fig. ~\ref{fig:quench}.

A large-grain hollow rod sample (12~cm long, 12~mm outer diameter, 8~mm inner diameter) was heat treated at 1400~\degree C for 3~h and the RRR obtained from the thermal conductivity measured at 4.2~K, $\kappa (4.2~K)$, [$RRR\approx4\times\kappa (4.2~K)$] decreased from 208 $\pm$ 12 to 152 $\pm$ 8 after HT. There was no change in the critical temperature after HT: $T_c = 9.26 \pm 0.01$~K before HT and $T_c = 9.24 \pm 0.02$~K after HT.

\subsection{\label{subsecA}Data analysis}
\begin{figure}
\includegraphics{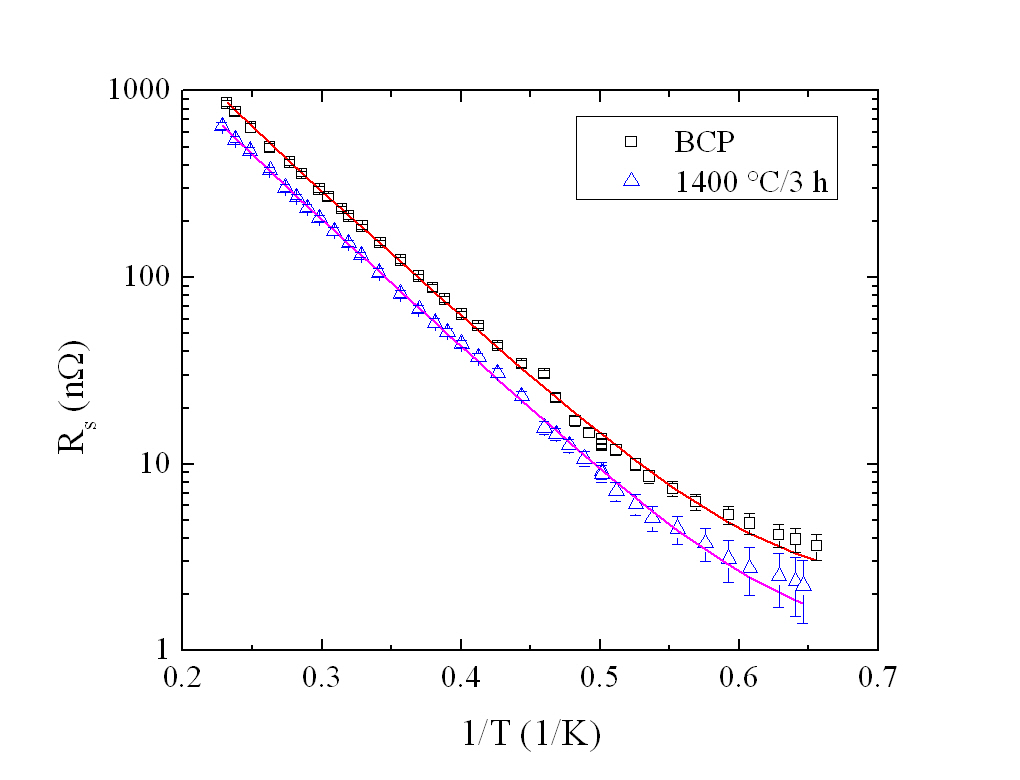}
\caption{\label{fig:rst} $R_s$ vs. $1/T$ measured after BCP and after HT at 1400~\degree C. Solid lines are least-square fits with $R_s(T) = R_{BCS}(T) + R_{res}$. The values of the fit parameters are $\Delta/kT_c=1.87\pm 0.02$, $\ell = (303\pm 85)$~nm, $R_{res}=(2.0\pm 0.3)$~n$\Omega$ after BCP and $\Delta/kT_c=1.90\pm 0.01$, $\ell = (76\pm 17)$~nm, $R_{res}=(1.0\pm 0.2)$~n$\Omega$ after HT at 1400~\degree C.}
\end{figure}

The low-field surface resistance as a function of temperature, $R_s(T)$, obtained from the ratio $R_s(T) = G/$\qz $(T)$, from each test has been fitted with a code \cite{Halb1970, Ciov2005} which calculates $R_{BCS}(T)$ to obtain material parameters such as the ratio $\Delta/kT_c$, where $\Delta$ is the energy gap value at 0~K, \textit{k} is Boltzmann's constant and $T_c$ is the critical temperature, and the normal electrons' mean free path, $\ell$. $T_c$, the London penetration depth ($\lambda_L$ = 32~nm) and the coherence length ($\xi_0$ = 39~nm) are considered material constants of Nb. The fit of $R_s(T)$ also yields the temperature-independent $R_{res}$. Figure~\ref{fig:rst}
shows, as an example, $R_s(T)$ data measured after BCP and after HT at 1400~\degree C and the curve fits with $R_s(T) = R_{BCS}(T) + R_{res}$.

\begin{figure}
\includegraphics{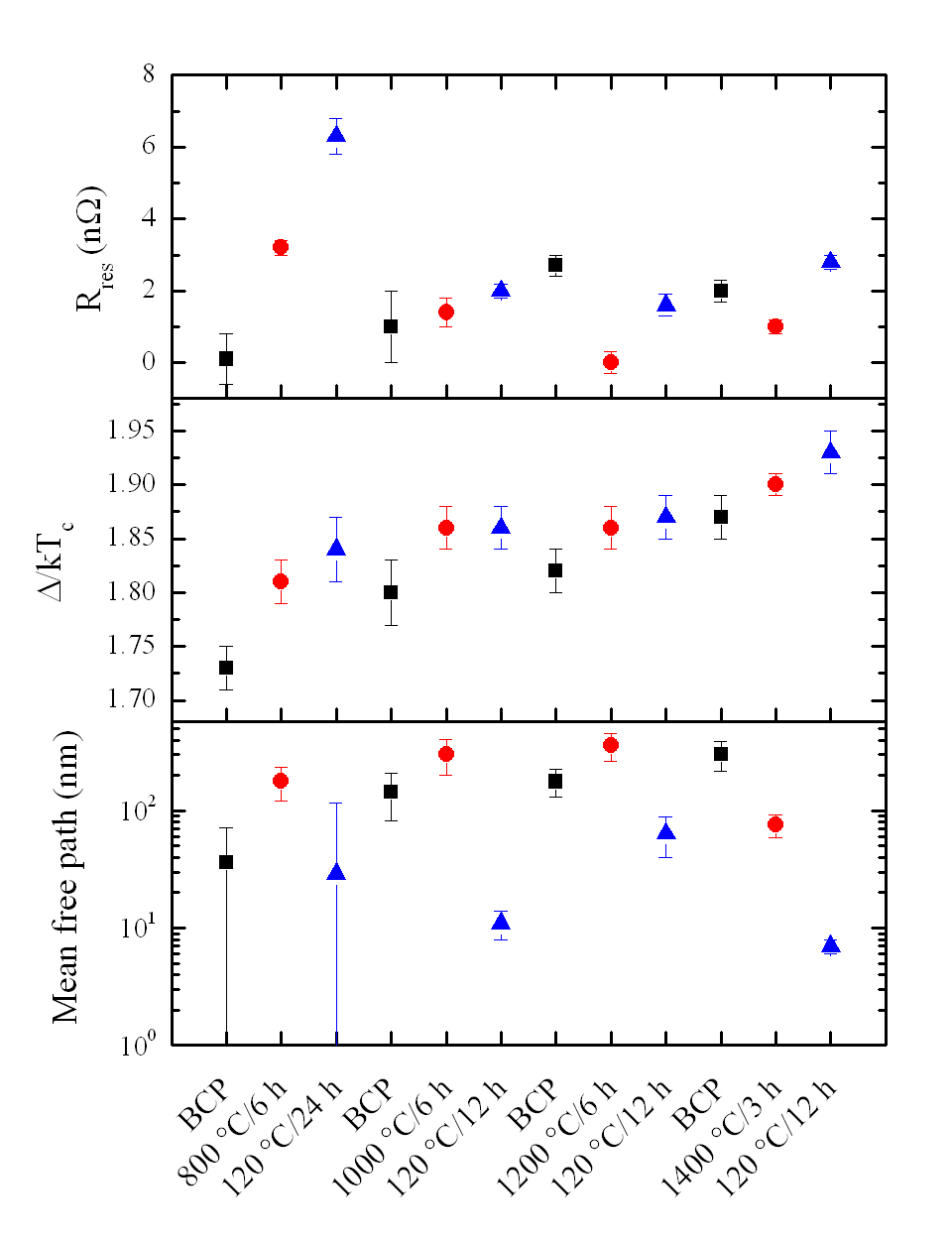}
\caption{\label{fig:ell} Mean free path, $\Delta/kT_c$ and $R_{res}$ from $R_s(T)$ fits for the various treatments: BCP (squares), HT (circles) and LTB (triangles).}
\end{figure}

Figure ~\ref{fig:ell}
shows $\ell$,  $\Delta/kT_c$ and $R_{res}$ obtained for the different treatments. Although the uncertainty in the value of $\ell$ is significant, there seems to be a pattern where HT increases the mean free path, except after HT at 1400~\degree C, whereas LTB reduces it. $\Delta/kT_c$ increases by $\sim12\%$ from the first BCP etching after all the treatments: it increases after HT and, less markedly, after LTB. The $R_{res}$ always increases after LTB. These dependencies will be discussed in relation to surface impurities in Sec. \ref{sec6}. Changes in $\ell$,  $\Delta/kT_c$ and $R_{res}$ by LTB are consistent with results from earlier studies \cite{Ciovati2004}.

\begin{figure}
\includegraphics{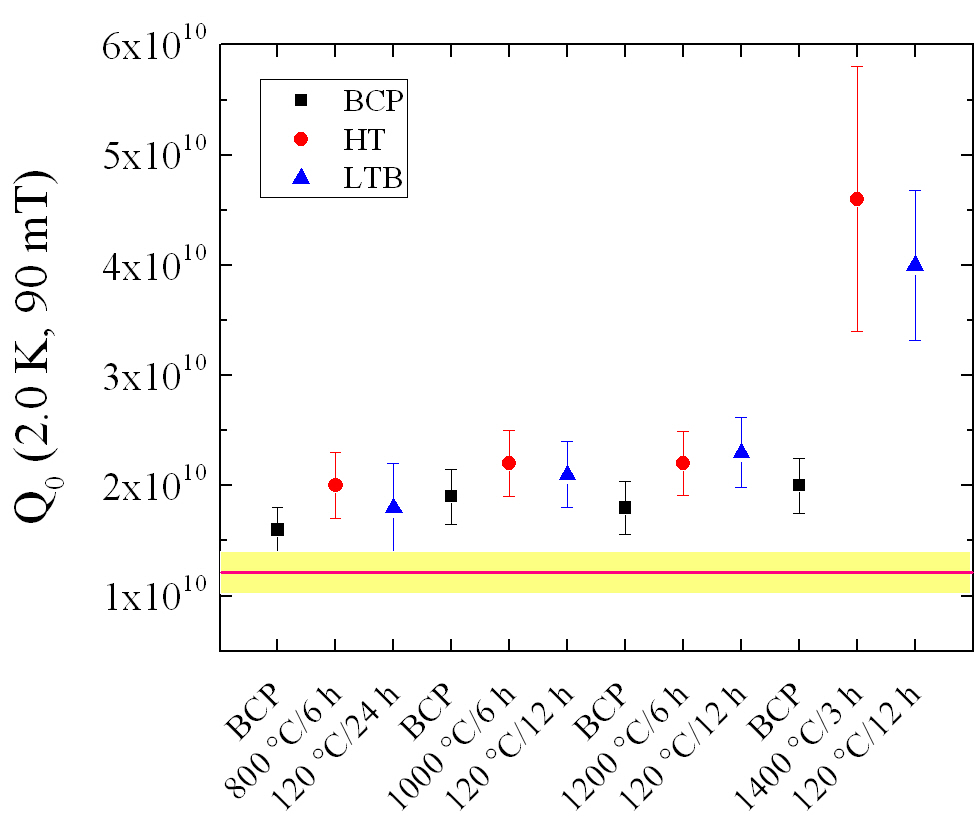}
\caption{\label{fig:Qplot} \qz(2.0 K, 90 mT) obtained after different treatments. The solid line is the average \qz(2.0~K, 70~mT) measured on fine-grain, electropolished 7-cell cavities for the CEBAF Upgrade. The yellow area indicates the standard deviation of the average.}
\end{figure}

Figure~\ref{fig:Qplot}
shows \qz(2.0 K, 90 mT) for the different treatments. The average \qz(2.0~K, 70~mT) = (1.2 $\pm$ 0.2)$\times 10^{10}$ measured in vertical test of 1.5 GHz 7-cell cavities for the CEBAF Upgrade is also shown in Fig. \ref{fig:Qplot} for comparison. The 7-cell cavities were built from standard high-purity, fine-grain Nb and treated by EP. Details about the 7-cell cavities preparation can be found in \cite{Reilly2011}. The data shown in Fig. \ref{fig:Qplot} indicate a high \qz\ value already after BCP and \qz\ improves by $\sim$20$\%$ on average after HT in the range 800-1200~\degree C. The LTB has no significant benefit on the \qz-value at 90~mT. As mentioned before, \qz\ more than doubles from the baseline value after HT at 1400~\degree C.

\section{\label{sec4}Results on samples}
Samples (7.5 mm $\times$ 5 mm $\times$ 3.125 mm in dimensions) were cut by wire electrodischarge machining from a niobium ingot. The samples were etched by BCP 1:1:1, removing $\sim$70~$\mu$m, heat treated in a UHV furnace at 600~\degree C for 10~h to degas hydrogen, and etched by BCP 1:1:2, removing $\sim$30~$\mu$m. Afterwards, the samples were nanopolished at Wah Chang, USA, to obtain a surface with mirror quality smoothness. Samples labeled L40 and L46 were heat treated with the cavity at 1000~\degree C. Sample L48 was heat treated with the cavity at 1200~\degree C, while samples L50 and L51 were heat treated with the cavity at 1400~\degree C. Sample L50 had an additional chemical etching by BCP 1:1:1 removing $\sim$80~$\mu$m after nanopolishing, prior to heat treatment. The samples labeled L11 and L35 were not heat treated after nanopolishing and were used as a reference for comparison with heat treated samples.

\begin{figure}
\includegraphics{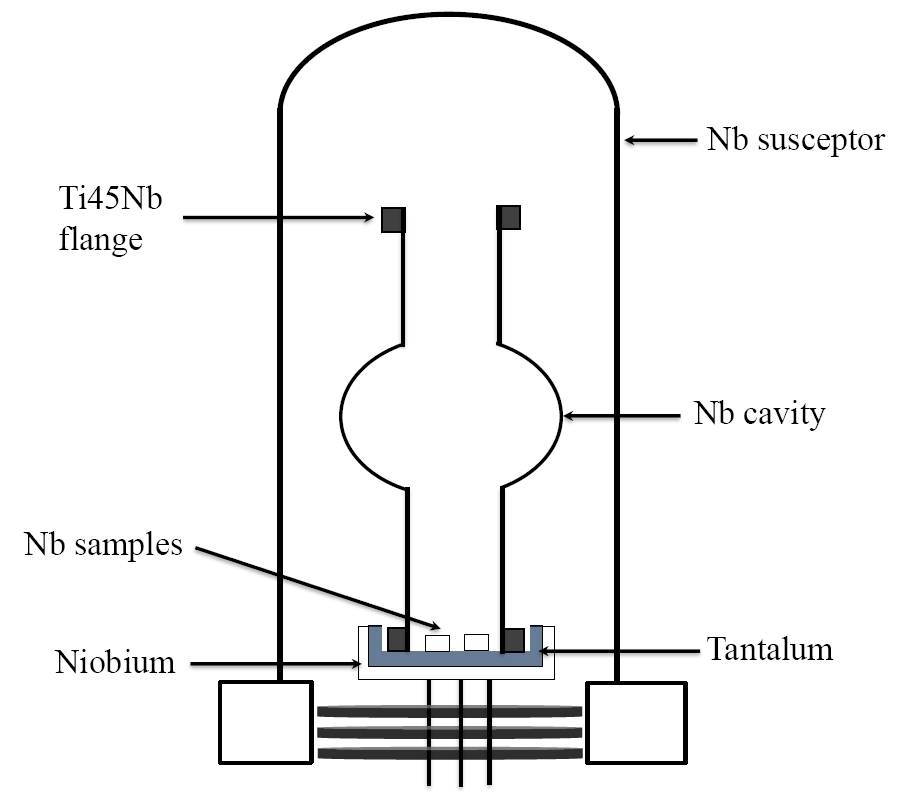}
\caption{\label{fig:sampleandcav} Schematic of the cavity and samples inside the furnace.}
\end{figure}

The samples' treatments attempted to replicate the cavity treatments to the extent possible, the differences being in the samples not being subjected to the deep-drawing process, being chemically-mechanically polished ("nanopolishing") instead of CBP, and each being heat-treated with the cavity only one time, instead of multiple BCP-HT cycles as for the cavity.
A schematic representation of the cavity and samples inside the furnace is shown in Fig.~\ref{fig:sampleandcav}.

\subsection{\label{subsecA}SIMS analysis}
Samples L40, L48 and L51 have been analyzed using a CAMECA IMS-6f magnetic sector dynamic SIMS to measure the depth profiles of H, C, N and O. Depth profiles for Ti were also measured on samples L48 and L51. For detection of H, C, O, N,  Cs$^+$ primary ion beam was used since Cs enhances the negative ion yields. Experiments were carried out using 14.5 keV impact energy, with a 120 $\times$ 120~$\mu$m$^2$ raster and a 30~$\mu$m diameter detection area to minimize contributions from crater edges and achieve good depth resolution. A current of 20~nA was maintained throughout the analysis for acceptable sputtering rates to reach depths of the order of 1-3~$\mu$m. Since absolute quantification of H in Nb is not possible using SIMS because of the high diffusion coefficient of H in Nb, H levels were characterized as H/Nb ratios. 
For metallic impurities, analysis was performed using 5.5~keV O$_2^+$ primary ion beam for better positive secondary ion yield. A 200~$\mu$m raster with a 60~$\mu$m detection area and a 120~nA primary beam current was used for this analysis.
A mass resolution, $M/\Delta M$, of 2050 was chosen so as to eliminate any interfering ions and improve analysis detection limits for measurements with Cs and O beams. The analysis chamber was also kept under UHV conditions ($\sim10^{-10}$~Torr) to minimize contamination.
Ion implantation was used to obtain standards of C, O and N in Nb \cite{Prateek2011}. Since N does not have a significant negative secondary ion yield in SIMS, NbN$^-$ ions were used to monitor the N$^-$ signal.

\begin{figure}
\includegraphics{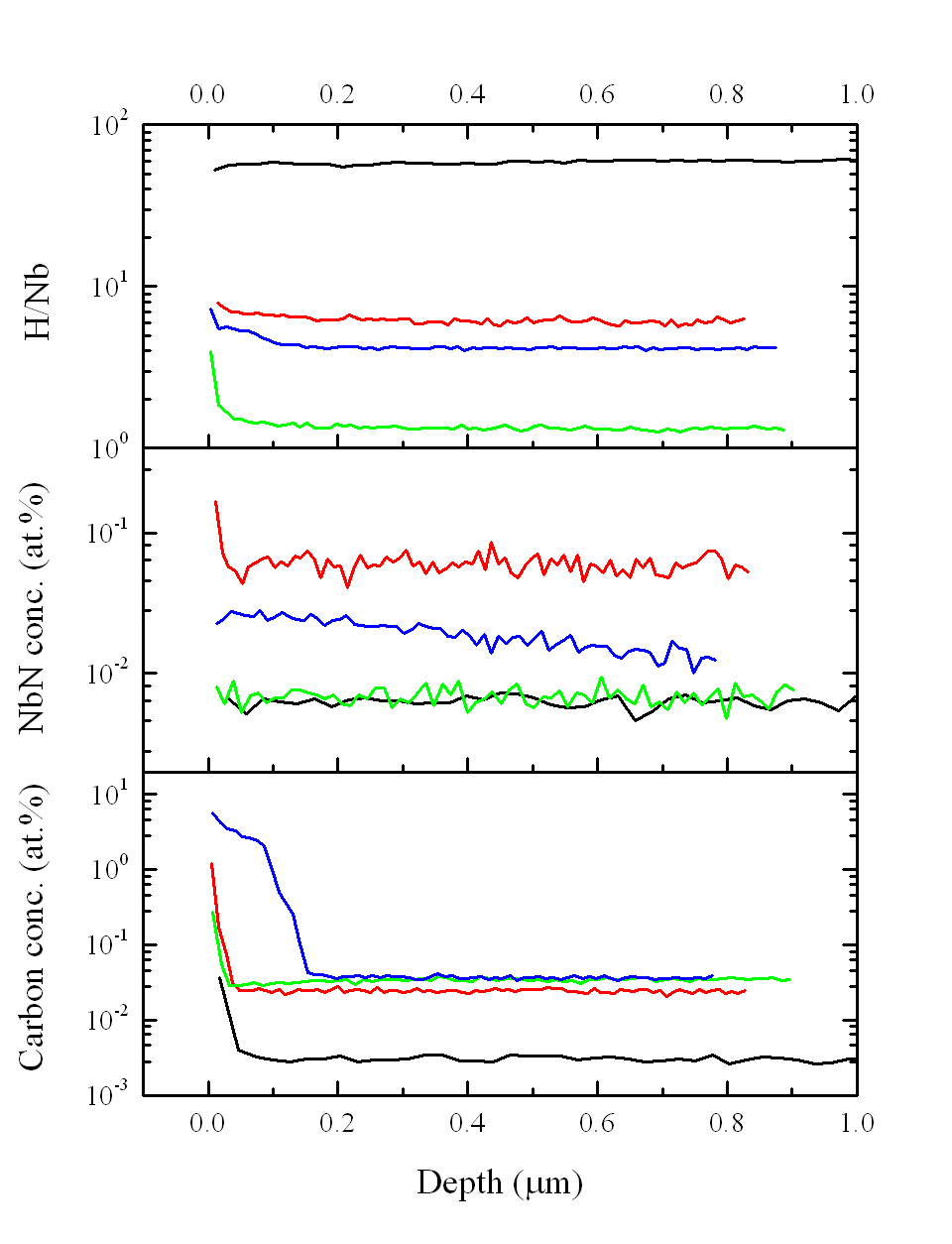}
\caption{\label{fig:SIMShigh1} H/Nb ratio, NbN and carbon concentrations measured in the reference, non heat-treated, sample L35 (black line), sample L40 heat treated at 1000~\degree C (red line), sample L48 heat treated at 1200~\degree C (green line) and sample L51 heat treated at 1400~\degree C (blue line).}
\end{figure}

\begin{figure}
\includegraphics{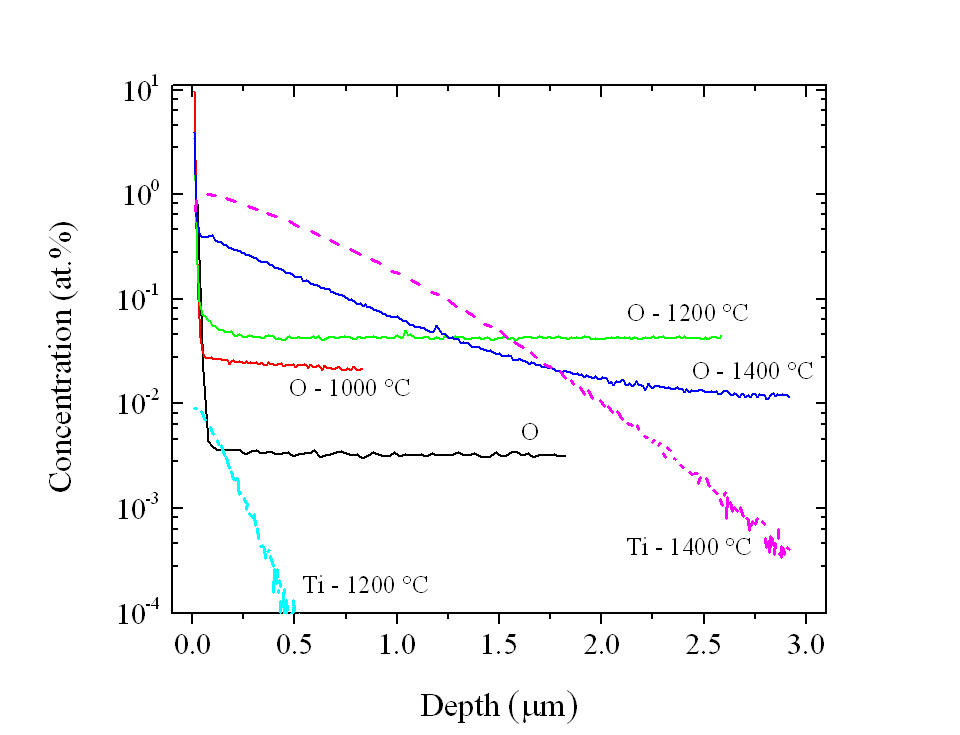}
\caption{\label{fig:SIMShigh2} Oxygen concentration measured in the reference, non heat-treated, sample L35 (black line), sample L40 heat treated at 1000~\degree C (red line), sample L48 heat treated at 1200~\degree C (green line) and sample L51 heat treated at 1400~\degree C (blue line). Also shown is the titanium concentration measured in samples L48 (cyan dahed line) and L51 (magenta dashed line).}
\end{figure}

Figure ~\ref{fig:SIMShigh1} 
shows the depth profiles of nitrogen, hydrogen and carbon measured in samples L35, L40, L48 and L51, whereas the depth profiles of oxygen and titanium in the same samples are shown in Fig.~\ref{fig:SIMShigh2}. Quantification of titanium was established by analysis of ion implanted $^{48}$Ti into Nb. The concentration of titanium in the reference sample was below the detection limit ($\sim$1~at.ppm).
Significant carbon and oxygen segregation near the surface resulted after the heat treatment at 1400~\degree C. A titanium concentration of $\sim$1~at.$\%$ near the surface was also found after 1400~\degree C. It was realized that Ti evaporated from the cavity flanges which are made of Ti45Nb. The vapor pressure of Ti at 1000~\degree C, 1200~\degree C and 1400~\degree C is $3.0\times 10^{-9}$~Torr, $7.6\times 10^{-7}$~Torr and $6.3\times 10^{-5}$~Torr, respectively \cite{Desai1987}.

\begin{figure}
\includegraphics{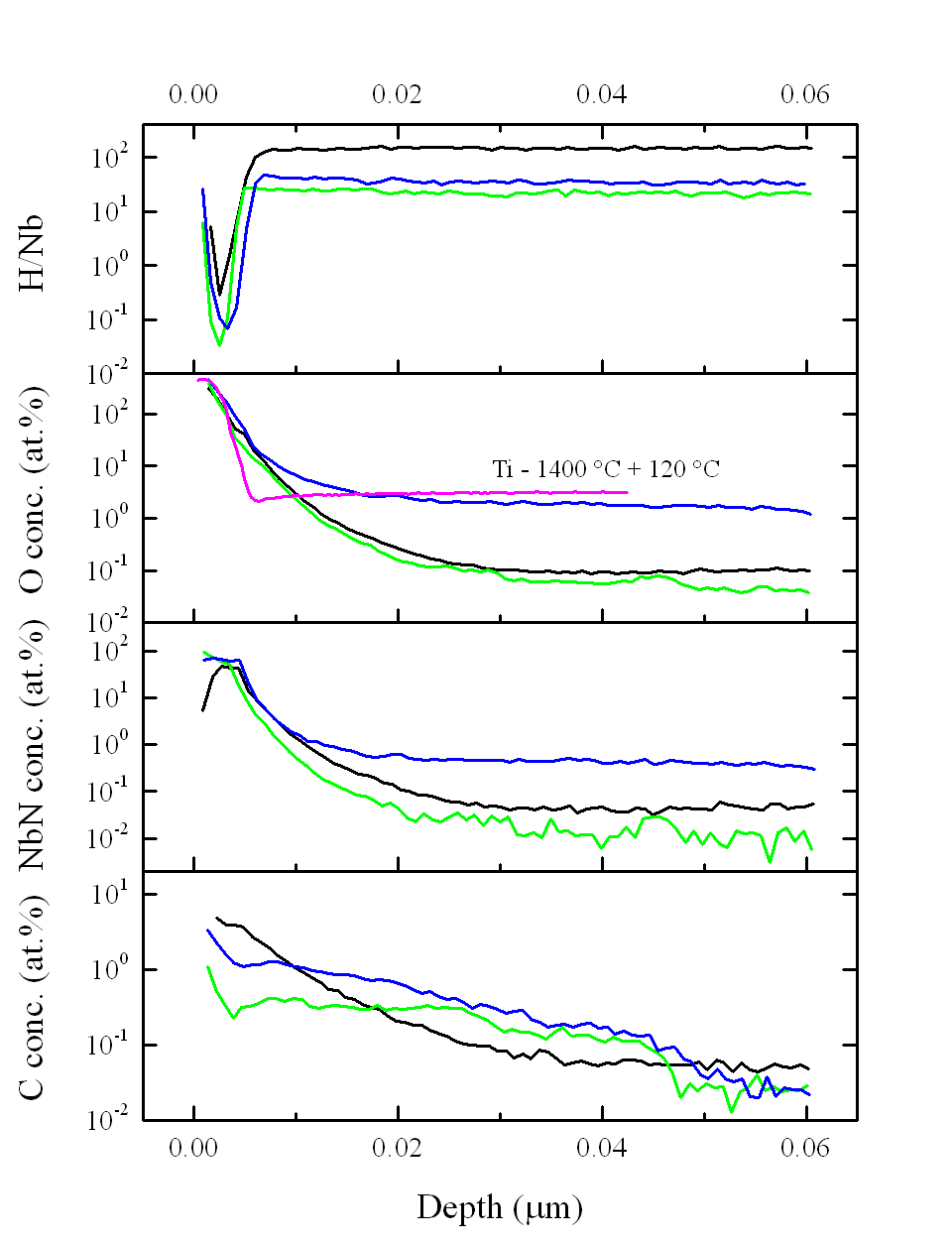}
\caption{\label{fig:SIMSlow} High-resolution depth profiles of H/Nb, oxygen, NbN, and carbon measured in a reference, non heat-treated, sample L11 (black line), sample L48 heat treated at 1200~\degree C (green line) and sample L51 heat treated at 1400~\degree C and 120~\degree C (blue line). Also shown in the second panel is the Ti depth profile in sample L51 (magenta line).}
\end{figure}

Since the penetration depth of the rf field is of the order of 40~nm, it is important to measure impurities depth profiles within $\sim$100~nm depth from the surface with $\sim$1-2~nm depth resolution. This can be accomplished by reducing the SIMS primary ion beam energy to 6~keV for Cs$^+$ beam and to 1.25~keV for O$_2^+$ beam. A primary ion current of 7~nA, a raster area of 200~$\mu$m $\times$ 200~$\mu$m with 60~$\mu$m detected area and a mass resolution of 2000 were used for Cs$^+$ beam, whereas a current of 20~nA, a raster area of 220~$\mu$m $\times$ 220~$\mu$m with 60~$\mu$m detected area and a mass resolution of 2200 were used for O$_2^+$ beam. SIMS measurements were repeated on another reference sample, labeled L11, on sample L48 and on sample L51 after it had been baked in UHV at 120~\degree C for 12~h. Figure~\ref{fig:SIMSlow} shows the results of the low-energy SIMS analysis on these samples. In the region below the surface oxide, the H level is significantly lower in the heat treated samples compared with the reference, non heat-treated, sample.

\subsection{\label{subsecB}PCT, EDX and XPS results}
Superconducting properties of the surfaces of samples L46 (heat treated at 1000~\degree C) and L50 (heat treated at 1400~\degree C) were analyzed through the use of electron tunneling spectroscopy. By cooling the sample below the critical temperature down to 1.8~K, tunneling measurements can directly probe the superconducting density of states, $N(\epsilon)$. Using a point-contact technique, we measured both the conductance and differential conductance spectra from which we extracted the superconducting gap of both samples, as shown for example in Ref.~\cite{Proslier2008}. Assuming the niobium sample is in the clean limit, the probing depth of the PCT measurement can be estimated to be on the order of the niobium coherence length ~40 nm.

In the setup at Argonne National Lab, the point-contact is made by approaching the sample surface with a sharpened gold tip to create a SIN (superconducting-insulating-normal) junction, where the natural oxide layer on the Nb surface creates the insulating layer. The tip-sample separation is mechanically controlled \textit{in situ} through the use of a differential screw while mechanical hysteresis allows for probing slightly different regions to obtain statistics on the sample's superconducting properties. A homemade analog sweep circuit combined with a lock-in amplifier were used to simultaneously measure the junction's $IV$ and $dI/dV\propto N(\epsilon)$ curves, respectively. The resulting spectra were normalized to the normal state differential conductance $dI/dV_{N}$ (assumed to be linear) and fit using the modified Blonder-Tinkham-Klapwijk (BTK) theory \cite{Dynes1978,BTK1982} to extract the gap $\Delta$, barrier strength $Z$, and the phenomenological quasi-particle lifetime broadening parameter $\Gamma$.

\begin{figure}
\includegraphics[width=0.5\textwidth]{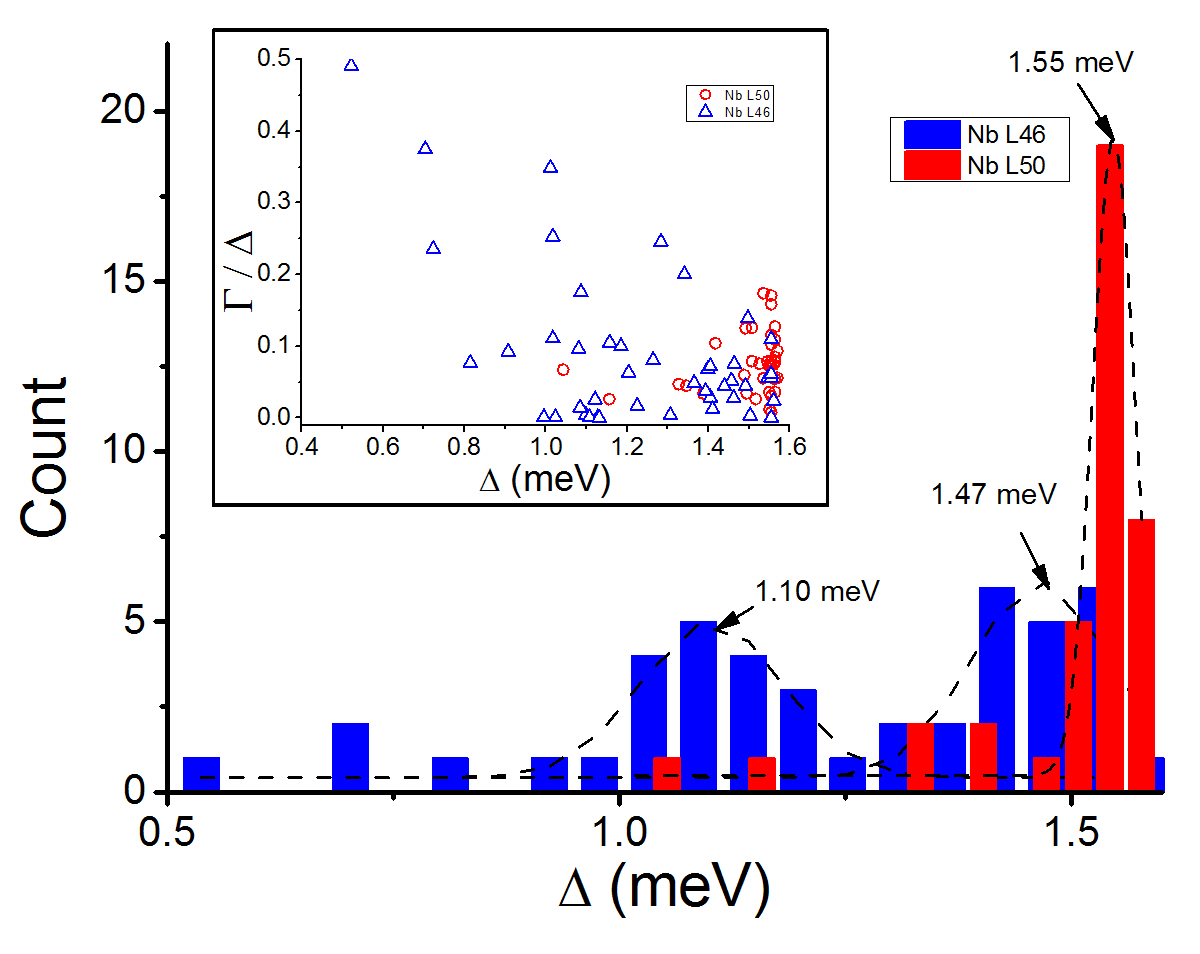}
\caption{\label{fig:junction_stats} A histogram of BTK fit parameter $\Delta$ for junctions measured by PCT spectroscopy. Sample L50 (red), heat treated at 1400 \degree C, shows a single sharp peak at the of 1.55~$\pm~$0.05~meV while sample L46 (blue), heat treated at 1000 \degree C shows two broad peaks at 1.10~$\pm~$0.16 and 1.47~$\pm~$0.15~meV. The dashed lines are gaussian fits to the histogram peaks where the FWHM defines the error in gap value. The ratio of $\Gamma/\Delta$ (inset) remains roughly constant for L50 (red circles) while there is a large deviation as gap values get smaller than about 1.4~meV for L46 (blue triangles).}
\end{figure}

\begin{figure}
\includegraphics[width=.5\textwidth]{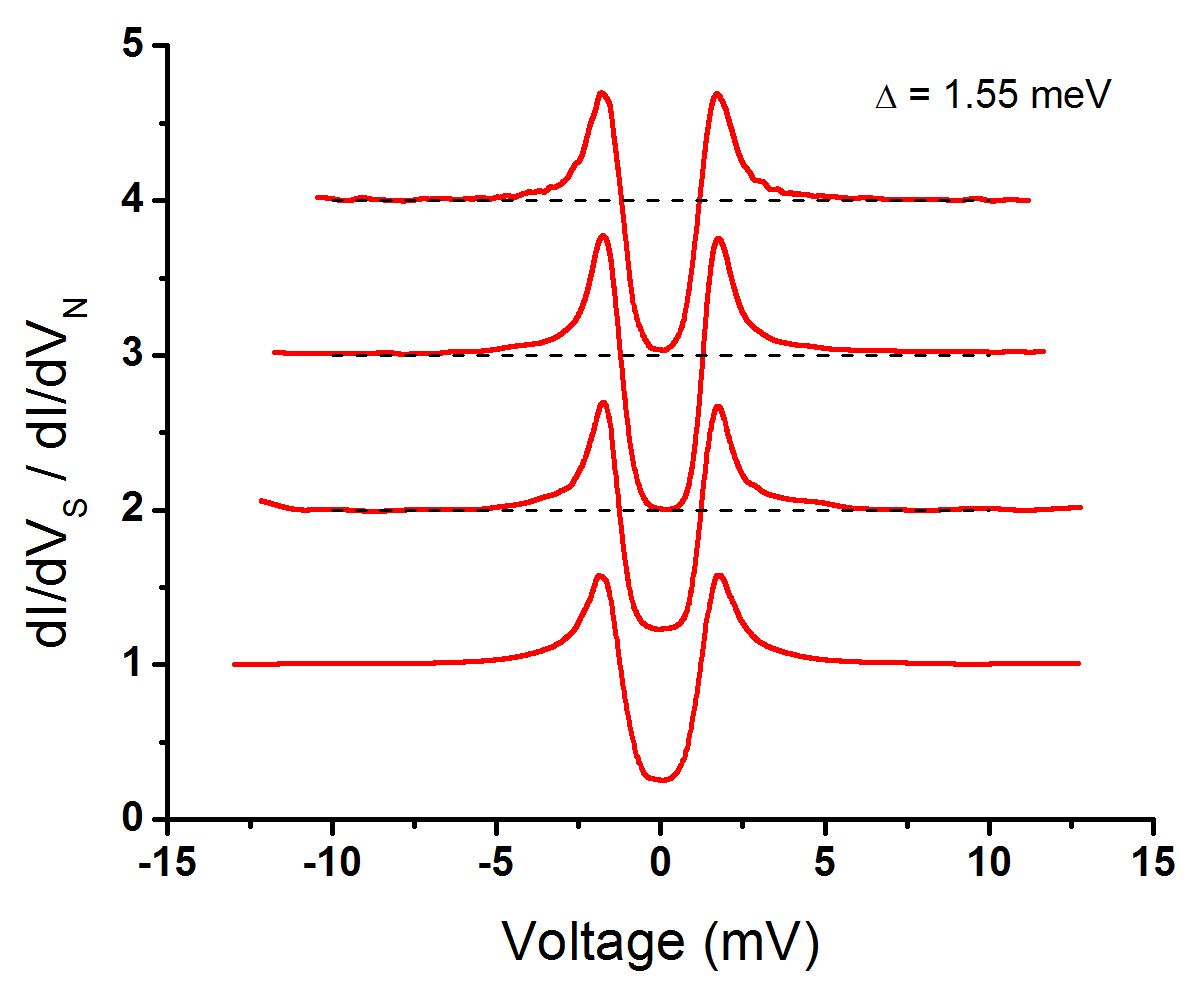}
\caption{\label{fig:L50SIN} Typical tunneling spectra obtained for different junctions on sample L50 offset vertically for clarity. The junctions obtained in L50 showed a consistent gap value of 1.55~meV and large coherence peaks (small $\Gamma$). }
\end{figure}

Approximately 40-50 junctions were measured on each sample to obtain a reasonable trend in the statistics on the superconducting gap. As shown by the red bars in Fig.~\ref{fig:junction_stats}, sample L50 exhibited a single gap sharply peaked at 1.55~meV, consistent with the bulk value for niobium. The spectra also displayed tall coherence peaks (small $\Gamma$) with an average $\Gamma$ value of 0.1~meV where the typical spectra obtained for L50 are shown in Fig.~\ref{fig:L50SIN}. The ratio of $\Gamma/\Delta$ was small and roughly constant for all the junctions measured for sample L50, which is a strong indication of a highly pure Nb and uniform surface with minimal inelastic scattering processes.

\begin{figure}
\includegraphics[width=.5\textwidth]{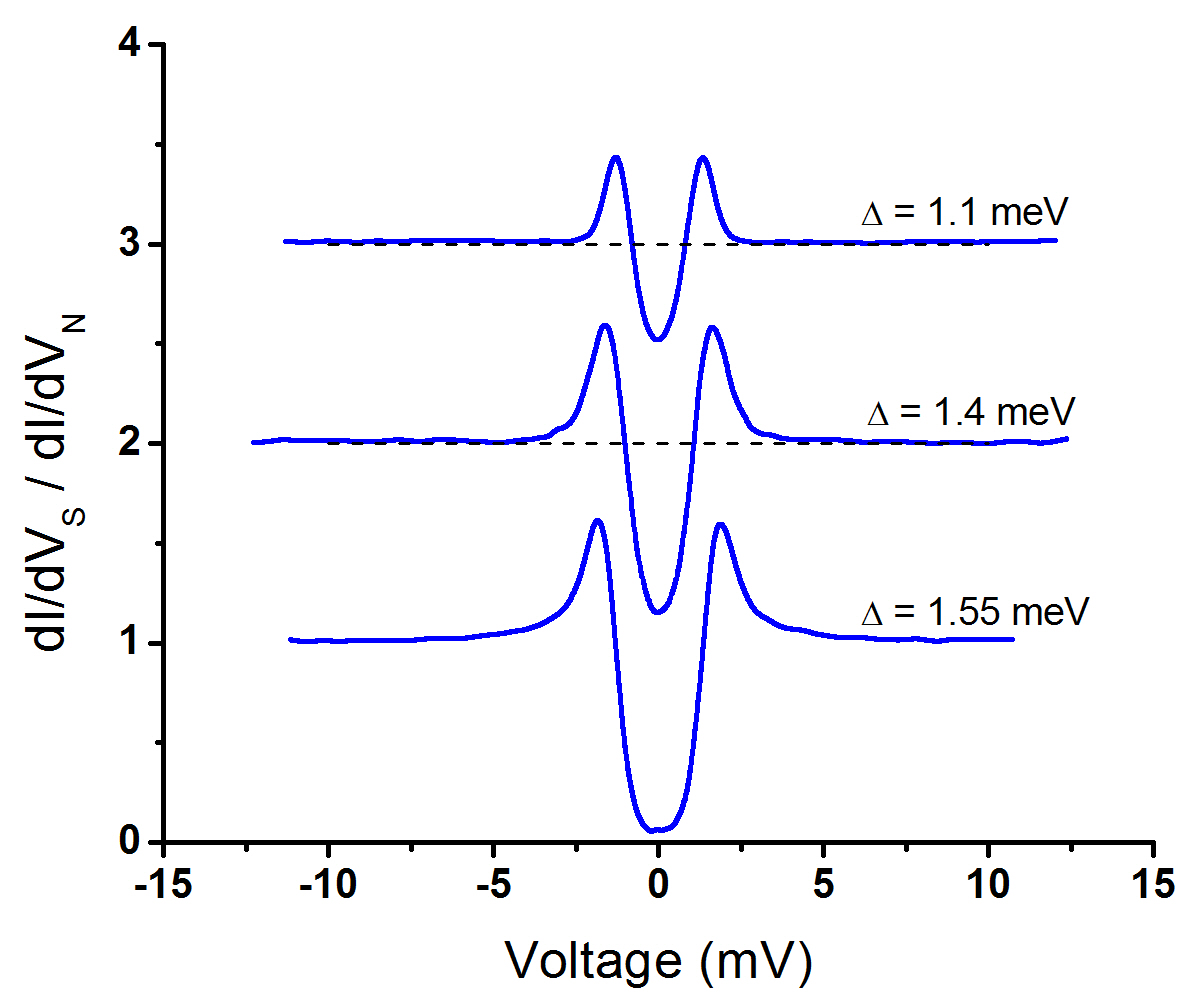}
\caption{\label{fig:L46SIN} Typical tunneling spectra obtained for different junctions on sample L46 offset vertically for clarity. This sample had a broad range of gap values with values of 1.10 and 1.47~meV being the most common. Junctions with gaps closer to the bulk value for Nb showed large coherence peaks (small $\Gamma$). However, this decreased considerably for gap values smaller than about 1.4~meV.}
\end{figure}

In contrast to the aforementioned, sample L46 shows the emergence of two different gap values of 1.10 and 1.47~meV (blue bars in Fig.~\ref{fig:junction_stats}), an indication of non-uniform superconducting properties of the surface possibly due to impurities. For the gap values near that of bulk Nb, $\Gamma$ is small and consistent with that of L50. However, as $\Delta$ gets smaller $\Gamma$ becomes quite large with significant variation. This can be seen in the spectra by lower and broader quasiparticle peaks along with higher zero bias conductance values in the differential conductance spectra, which are characteristic signatures of inelastic scattering processes. The typical differential conductance curves obtained for L46 are shown in Fig.~\ref{fig:L46SIN} where the gap size is increasing from 1.1~meV in the top curve to 1.55~meV in the bottom curve. Notice the peaks (and dip) features are getting more pronounced as the gap gets closer to that of bulk niobium. 

\begin{figure}
\includegraphics[width=.5\textwidth]{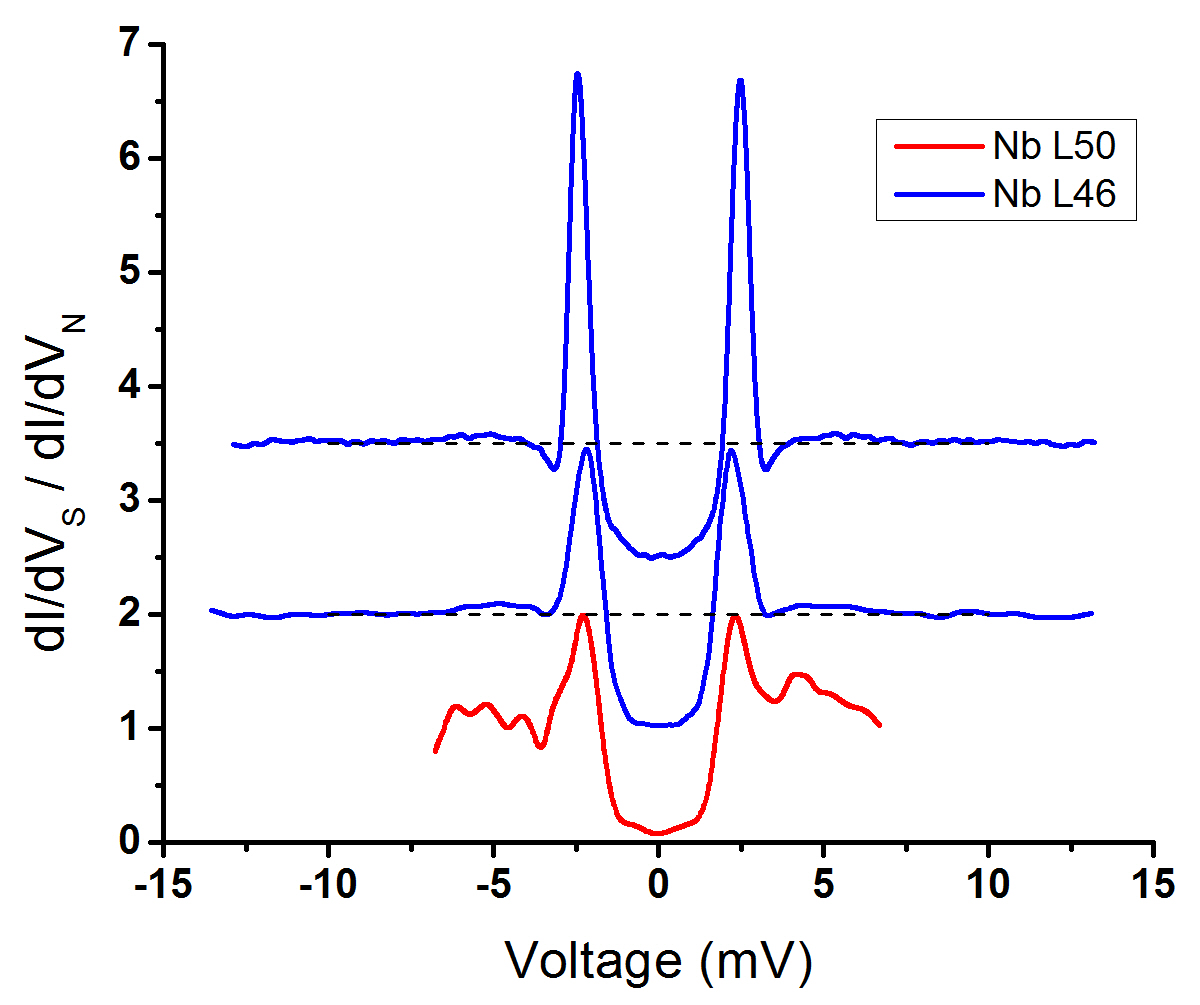}
\caption{Some SIS junctions obtained for samples L46 (blue) and L50 (red). The dip features seen outside the gap indicate a proximity effect of the Nb in the Au tip.}
\label{SIS}
\end{figure}

Niobium naturally forms a thin oxide layer when exposed to oxygen or water, creating a tunnel barrier. However,  two peculiar behaviors were noticed during the measurement of both samples as compared to previously measured cavity grade Nb samples under identical conditions \cite{Proslier2008}. Firstly, measurable tunneling junctions (resistance less than or equal to $10^{6}~\Omega$) were found after an unsually high number of turns during the mechanical approach (effectively as if the distance sample to tip was larger). The shape of the Au tips observed under an optical microscope after the measurements revealed a flattened apex area larger than usually seen. It is therefore likely that the native oxide thickness or/and workfunction is higher than previously measured on Nb samples. Secondly, several  SIS (superconductor-insulator-superconductor) junctions rather than SIN were measured on both samples, indicating that the Au tip dislodged a Nb piece from the sample's surface. In particular, this was more pronounced in sample L46 where six SIS junctions were found as opposed to the two found in L50. These results seem to point to a more brittle surface on that of L46 and may also explain the two gap trend and larger $\Gamma$ values. Some of the differential conductance spectra for the SIS junctions found in both samples are shown in Figure~\ref{SIS}. The dip in the SIS spectra around 3~meV for L46 (blue curves in Fig.~\ref{SIS}) are an indication of a proximity effect in the Au tip, meaning the dislodged Nb was in good metallic contact with the Au tip. It is noteworthy to point out that, although suprising, the number of SIS junctions obtained on both samples represent only up to 10\% of the SIN junctions measured. Therefore it is very unlikely that the trend in the superconducting gap and gamma parameters observed and discussed earlier could have been induced by the pressure exerted by the tip onto the samples.

Assuming the same value of $T_c = 9.25$~K as used for the computation of $R_{BCS}$ to fit of the $R_s(T)$ data, the gap values measured by PCT correspond to a ratio $\Delta/kT_c$ of 1.94 for sample L50 and of 1.84 for sample L46. These values are in very good agreement (less than 2\% difference) with those obtained from the fit of $R_s(T)$, shown in Fig. ~\ref{fig:ell}, even though the cavity was not subjected to the same nanopolishing surface treatement as the measured coupons. In addition,  the tunneling spectroscopy results showing that the 1400~\degree C HT improved significantly the surface superconductivity in terms of the amplitude of the gap and the pair-breaking parameter, as well as their uniformity, are consistent with the improvement of the cavity \qz-value.

\begin{figure}
\includegraphics[width=.5\textwidth]{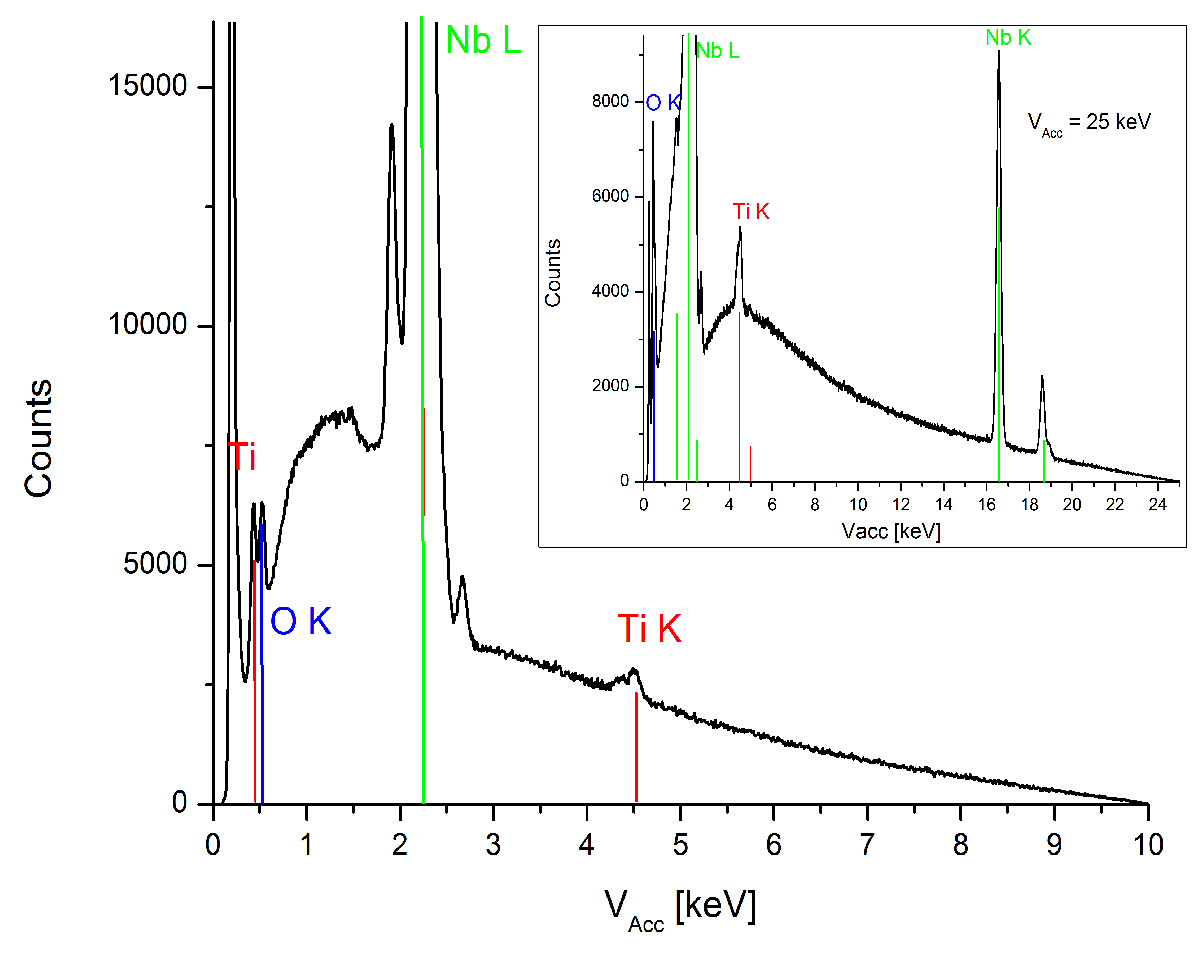}
\caption{Electron dispersive X-ray spectrum of sample L50 at 10 keV (main figure) and 25 keV (inset) show the presence of titanium (red lines) and oxygen (blue lines) on the surface of the niobium (green lines). The filament current was 10~$\mu$A, the magnification was 5000 and the aperture was 100~$\mu$m.}
\label{concEDX}
\end{figure}

\begin{figure}
\includegraphics[width=.5\textwidth]{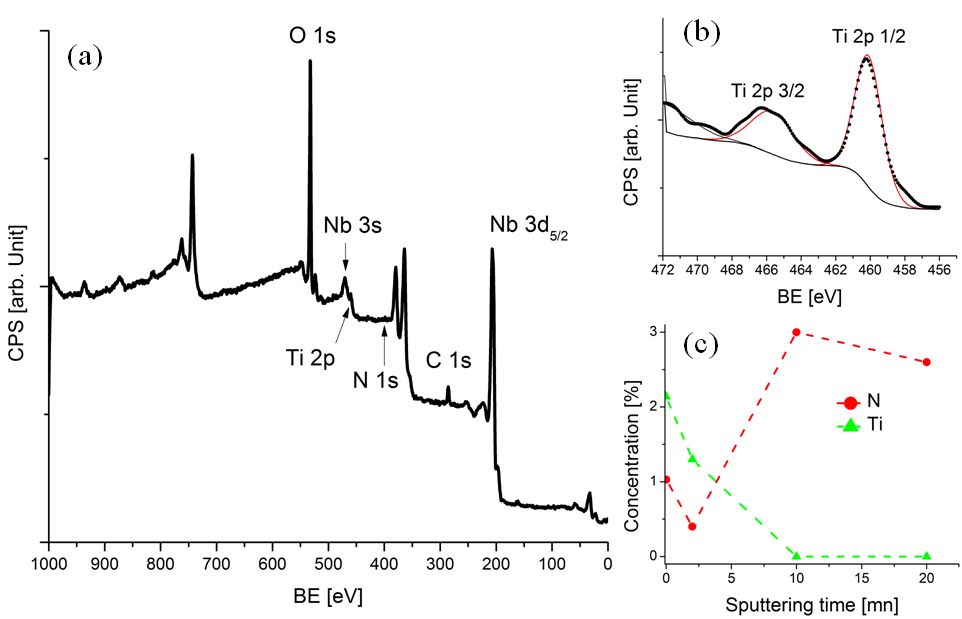}
\caption{The XPS spectrum of sample L50 (a) shows the niobium, titanium and oxygen peaks present on the surface prior to using the sample as the sputtering target. The upper right figure (b) shows zoomed in view of (a) around the Ti peaks where the red curve is the fit to the spectrum. The bottom right figure (c) shows the concentration of titanium (green) and nitrogen (red) measured by XPS as a function of sputtering time. After 10 minutes there is no longer any measurable titanium present, thus, the Ti impurities lie only on the surface of the sample.}
\label{TINconc}
\end{figure}

Following the PCT measurements, elemental analysis was performed on sample L50 to determine what impurities it might contain. First, using an SEM (Hitachi S4700) equipped with EDX, the chemical composition of the surface was probed at 10 and 25~keV (Fig.~\ref{concEDX}) showing the presence of a small amount of titanium as well as oxygen. The Ti concentration is estimated at $\sim$1~at.$\%$, close to the sensitivity of the instrument. To determine whether the Ti impurities resided on the surface or throughout the bulk Nb, sample L50 was used as a target in a sputtering system equipped with XPS. This was a home-built system that allowed a sample to be sputtered and then measured by XPS without breaking vacuum. The initial XPS spectrum is shown in Fig.~\ref{TINconc}(a) with a zoomed in view of the Ti peaks in Fig.~\ref{TINconc}(b). The surface was then sputtered several times at fixed time intervals with a chemical analysis at each interval. As shown in Fig.~\ref{TINconc}(c), the Ti concentration decreases with increasing sputtering time and after 10 minutes of sputtering no measurable Ti is seen in the XPS data. The results confirm the presence of Ti on the sample's surface, consistent with the results from SIMS.

\subsection{\label{subsecC}XRD results}
\begin{figure}
\includegraphics[width=80mm]{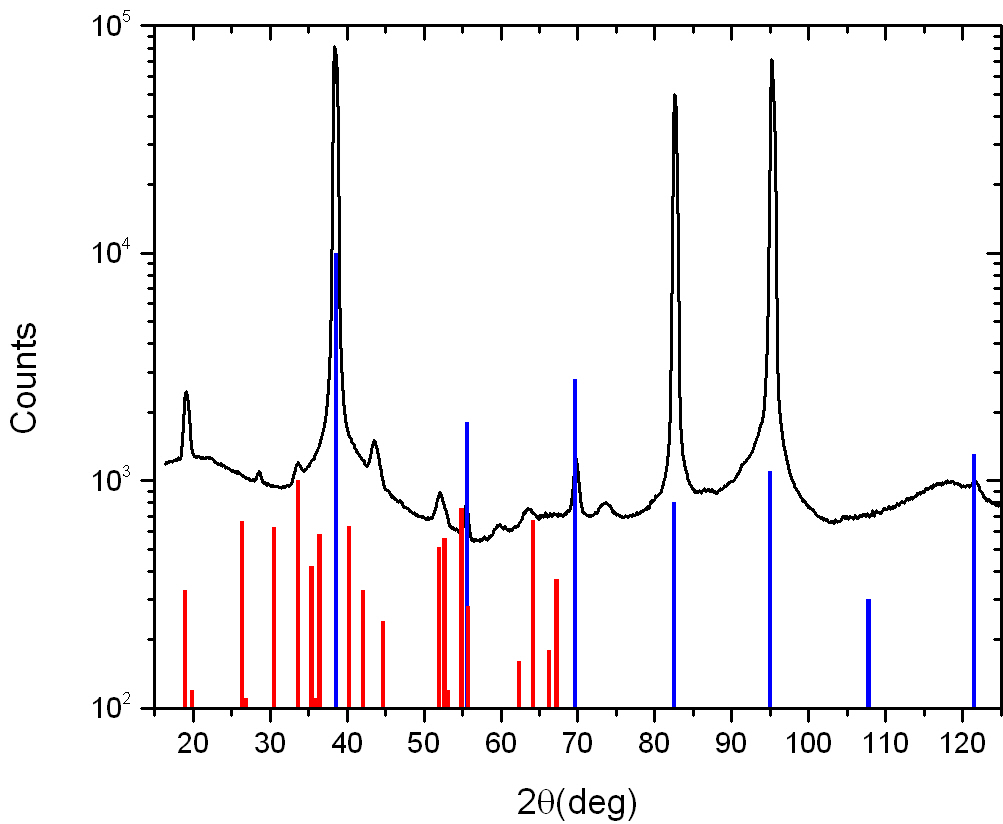}
\caption{XRD spectrum of sample L51. The vertical scale is expanded to show the presence of minor peaks. Vertical bars represent the relative intensities of tabulated poweder diffraction spectra for bulk Nb (blue) and $\gamma$-Ti$_3$O$_5$ (red).}
\label{XRD_WM}
\end{figure}

\begin{figure}
\includegraphics[width=80mm]{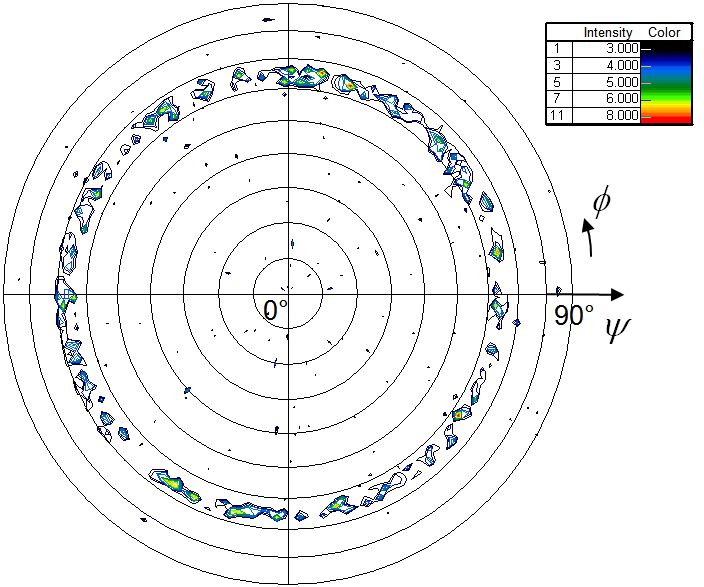}
\caption{Pole figure of (3 1 0) $\gamma$-Ti$_3$O$_5$ measured on sample L51. $\phi$ is the angle of rotation around the direction normal to the sample. $\psi$ is the tilt angle from the normal direction.}
\label{PF_Ti3O5}
\end{figure}

The XRD spectrum was measured for sample L51 with a Bruker SMART APEX II instrument at The College of William \& Mary. The instrument has a fixed-position source with a wavelength of 1.5406 \AA  (Cu-$K\alpha$ line) and a movable detector ($2\theta$) and sample ($\omega$). The detector is a charge coupled device (CCD) which covers about 30\degree in $2\theta$ per image. Four image positions were used for the experiment. In each case, the angle $2\theta$ is the center position of the 30\degree CCD image, and each $\omega$ value is set so that the angles of incidence and diffraction are equal. Long (10 min) exposure time was used in order to look for the presence of minor phases. Figure~\ref{XRD_WM} shows the detector counts as a function of $2\theta$. Besides the large peaks corresponding to bulk Nb orientations, smaller peaks are visible which showed some superficial match with $\gamma$-Ti$_3$O$_5$ or $\alpha$-NbN. To further investigate this, pole figures at fixed $2\theta$-values corresponding to high-intesity peaks for either $\gamma$-Ti$_3$O$_5$ or $\alpha$-NbN were measured using a PANalytical X'Pert PRO MRD instrument at Norfolk State University. Clear indication of a polycrystalline $\gamma$-Ti$_3$O$_5$ with fiber texture can be seen from the pole figure of (3 1 0)  $\gamma$-Ti$_3$O$_5$, shown in Fig.~\ref{PF_Ti3O5}. On the other hand, counts for (6 2 0) or (14 6 2) $\alpha$-NbN were within the noise.

\begin{figure}
\includegraphics[width=80mm]{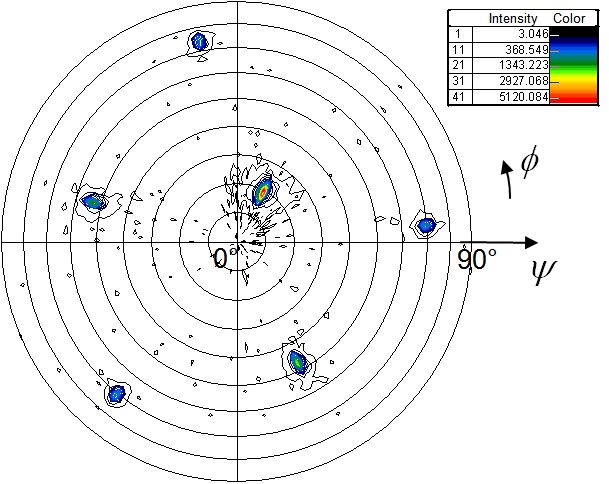}
\caption{Niobium (1 1 0) pole figure measured on sample L51.}
\label{PF_Nb}
\end{figure}

\begin{figure}
\includegraphics[width=80mm]{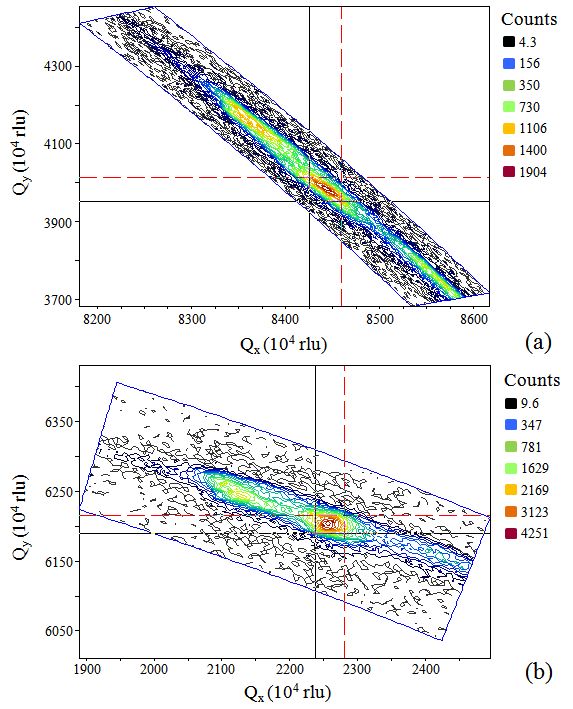}
\caption{Reciprocal space maps of Nb (1 1 0) pole measured on sample L51 with low (a) and high (b) X-ray angle of incidence.}
\label{RSM}
\end{figure}

The Nb (1 1 0) pole figure measured on sample L51 is shown in Fig.~\ref{PF_Nb}. The sample is a single crystal with (1 1 0) orientation with $\sim$20\degree inclination with respect to the normal to the sample's surface. A method which allows investigating the crystal quality consists of 2D-reciprocal space mapping (RSM) of an X-ray diffraction peak. In this case, the so-called triple axis mode \cite{Fewster1993} was used. This measurement was done at an incident angle of $\sim$4.9\degree and $\sim$21.9\degree, corresponding to a probing depth of $\sim$680~nm and $\sim$3~$\mu$m, respectively. By doing the measurements both at low and high incident angles, changes in crystal quality with increasing depth from the surface can be revealed. Reciprocal space maps obtained at low and high incident angle are shown in Fig.~\ref{RSM}. The full width half maximum (FWHM) of the peak in the reciprocal space maps is related to the crystal quality, the lower the FWHM, the better the crystal quality is. The relative variations, $\delta Q/Q$, of the reciprocal lattice vectors in the directions parallel ($Q_x$) and normal ($Q_y$) to the surface are directly related to variations in the spacing of the crystal plane and were measured to be $\delta Q_x/Q_x = 1.9\%$ and $\delta Q_y/Q_y = 0.4\%$ for high angle of incidence, $\delta Q_x/Q_x = 0.4\%$ and $\delta Q_y/Q_y = 1.6\%$ for low angle of incidence. There is no large variation of crystal quality with depth, although there seems to be opposite change between \textit{x} and \textit{y} directions.

\section{\label{sec5}Mechanical properties}
One of the issues related to the heat treatment of standard high-purity fine-grain Nb at temperatures greater than $\sim$1000~\degree C is a reduction of the yield strength due to re-crystallization during the HT \cite{Rao1994}. The JLab specifications for the yield strength and tensile strength of high-RRR Nb at room temperature are 48.2~MPa and 96.4~MPa, respectively. For the mechanical analysis of a multi-cell cavity, Von Mises stresses are calculated under a variety of load conditions, both at room temperature and 2.0~K, and it is verified that, in each condition, the peak stresses are less than 2/3 of the yield strength \cite{Cheng2010}.

A total of eight ASTM-compliant samples with nominal reduced section gage 45~mm long $\times$ 6~mm wide $\times$ 4~mm thick were cut by wire-EDM from the same ingot G used to fabricate the single-cell cavity. The samples were etched by BCP 1:1:1, removing $\sim$100~$\mu$m. Four samples, labelled No. 5-8, were heat treated at 1400~\degree C for 3~h with the same procedure used for the cavity HT. The tensile properties of all eight samples were measured at the National High Magnetic Field Laboratory according to ASTM E8 and ASTM E1450, on a servo-hydraulic test machine in displacement control. The test machine is equipped with a cryostat that enables low temperature tests with the specimen and test fixture immersed in sub-cooled ethanol (182~K), or liquid nitrogen (77~K), or liquid helium (4~K). Strain is measured with two 25 mm gage length extensometers mounted on opposite sides of the specimen and their outputs are averaged to compensate for inaccuracies that may occur due to machine tolerances or specimen misalignment. The final percent elongation is determined from lines that are scribed 25 mm apart in the gage section of the sample. At the start of a test, the initial displacement rate is 0.5~mm/min (elastic strain rate = 1.8$\times 10^{-4}$ s$^{-1}$)  and it is increased after the onset of yield (at about 1.5\% strain) to a rate of 1.0~mm/min. An unload/reload cycle is performed at 1.5\% strain to confirm the elastic modulus. 

\begin{figure}
\includegraphics[width=80mm]{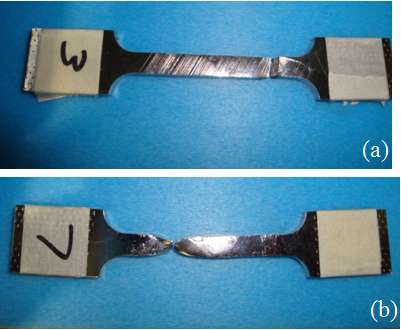}
\caption{\label{fig:Tensile} Images of sample No. 3 (a) and No. 7 (b) after tensile test at 77 K.}
\end{figure}

The test results are listed in Table~\ref{tab:tabl2}. The data show that the yield strength at 295 K actually increased after HT, with no significant change in Young's modulus and $<10\%$ reduction in tensile strength. The samples have only few grains in the gage section and typically only one grain in any local region of the cross-section.  The deformation of the large grains at low temperature resulted in interesting deformation of the tensile specimens. Examples of the non-uniform strain and unique deformation characteristics are shown in photos of fractured samples in Fig.~\ref{fig:Tensile}.

\begin{table*}
\caption{\label{tab:tabl2} Results from tensile tests (\textit{T}: temperature, \textit{E}: Young's modulus, \textit{YS}: yield strength at 0.2\% strain, \textit{TS}: tensile strength, $\epsilon_b$: elongation at fracture, $A_r$: area reduction) conducted at different temperatures on non heat-treated samples and samples heat treated at 1400~\degree C for 3 h.
}
\begin{ruledtabular}
\begin{tabular}{cccccccc}
 Sample No.&\textit{T} (K)&HT&\textit{E} (GPa)&\textit{YS} (MPa)&\textit{TS} (MPa)&$\epsilon_b$ (\%)&$A_r$(\%)\\ \hline
 1&295&None&72&43&88&44.0&99.0 \\
 2&183&None&76&225&234&18.3&29.6\\
 3&77&None&74&425&562&28.2&255.8\\
 4&4.2&None&86&$\cdot \cdot \cdot$&894&1.5&1.6\\
 5&295&1400 \degree C$/$3 h&73&56&82&21.8&99.0\\
 6&182&1400 \degree C$/$3 h&74&259&294&22.3&99.0\\
 7&77&1400 \degree C$/$3 h&103&407&509&26.0&99.0\\
 8&4.2&1400 \degree C$/$3 h&105&$\cdot \cdot \cdot$&563&7.2&2.7\\
 \end{tabular}
\end{ruledtabular}
\end{table*}

\section{\label{sec6}Discussion}
From a broad perspective, the cavity test results shown in Sec. \ref{sec3} clearly indicate the benefit of using ingot Nb and high-temperature HT to produce SRF cavities with exceptionally high \qz\ at accelerating gradients of $\sim$20-25~MV/m. Besides the performance improvement, the cavity material and processing used in this study is expected to provide significant cost savings compared to the standard Nb material and processes currently used for the production of SRF cavities.
SIMS analysis show that the hydrogen content is reduced by up to about one order of magnitude in samples which were heat treated at 800-1400~\degree C compared to non heat treated ones. Changes in the concentration of other impurities such as C, N, and O as a function of heat treatment temperature are less obvious, but generally increases after HT. The concentration of these impurities is of the order of $10^{-1} - 10^{-2}$~at.\%, significantly lower than that of hydrogen. The highest O, C and N concentrations in the near-surface region were found after HT at 1400~\degree C. Yet, parameters indicative of the quality of the superconducting state such as the energy gap, the quasi-particle lifetime broadening and the surface resistance all improved after this heat treatment. The residual resistance also decreased compared to that prior to the HT, after BCP. Furthermore, SIMS analysis, corroborated by data using EDX and XPS, indicates the presence of a diffusion profile for Ti in Nb, with a concentration of $\sim$1~at.\% at the surface, and for O, which is gettered by Ti during the HT at 1400~\degree C. This result was unintentional and explained by the cavity flanges being made of Ti45Nb. The improvement of superconducting properties after HT at 1400~\degree C seems couterintuitive, given the impurity analysis. 

It is not possible at this stage, and it is beyond the scope of this article, to provide a  conclusive explanation of the mechanism behind the extended increase in \qz($B_p$) which resulted in a very high \qz-value at 90 mT. Related to this phenomenon it should be noticed that: (i) a similar \qz-rise had been reported for a 3~GHz niobium cavity which had been oxidized \textit{in situ} with dry oxygen gas after HT at 1950~\degree C for 9 h \cite{Parodi1999}; (ii) the \qz-rise we've measured can be well described by a recent numerical calculation of the surface resistance based on the Mattis and Bardeen theory modified to account for moving Cooper pairs under the action of the rf field \cite{Binping2012}. For the dry oxidation treatment described in \cite{Parodi1999}, the cavity surface was exposed to 0.5~atm of pure oxygen for 10 days while in our experiments the pressure was 1 atm and the time was not longer than 1 h, corresponding to an exposure lower by about two orders of magnitude. Further studies are necessary to investigate the role of dry oxidation after heat treatment.

Further studies are also needed to understand the effect of the interplay among Ti, O and H on the rf superconducting properties of annealed Nb. Regarding this effect, it should be mentioned that the reduction of H concentration and the presence of Ti and O at the Nb surface after heat treatment at 1400~\degree C, acting as trapping centers for H \cite{Magerl1983,Cantelli2006}, both contribute to prevent the formation of normal-conducting hydrides, which had been found in the past to be responsible for an increase of the residual resistance \cite{Bonin1991, Isagawa1980, Knobl2002}.

The results from the tensile tests provided the important information that the mechanical properties of ingot Nb do not degrade after 1400~\degree C HT, which is important when considering this process for multi-cell cavities.

\section{\label{sec7}Conclusion}
The measurement results described in this contribution push the limit of high quality factors in SRF cavities at temperature and surface fields relevant for present and future CW accelerators relying on SRF cavity technology \cite{Harwood2001, Corlett2011, Nakamura2012}. These results have been accomplished by using ingot Nb material of medium purity with final treatment consisting of a high-temperature annealing in a clean induction furnace. The highest achieved \qz-value is about a factor of four higher than the average \qz-value achieved in the production cycle of multi-cell cavities for the CEBAF 12~GeV Upgrade, which used the standard "recipe" of fine-grain, high purity Nb with final treatment consisting of EP + LTB. It should also be remarked that the material and processes which resulted in the much higher \qz-value are less expensive than the current standard material and processes.
Unlike for fine-grain Nb, the mechanical properties of medium purity, ingot Nb did not degrade after 1400 \degree C annealing which makes this material/process combination suitable for multi-cell cavities.

The \qz($B_p$) dependence in the rf test which exhibited the highest \qz-value had an extended low-field Q-increase. An important question, from both the theoretical and experimental point of view, is how high can the peak \qz-value be pushed and also what is the highest surface field this peak \qz-value can be attained at.

A thorough analysis of Nb samples annealed with the cavity using several complementary surface analytical methods revealed increased concentrations of C, N, and O after annealing. The hydrogen, on the other hand, was significantly reduced. A titanium-oxide phase, within $\sim$1~$\mu$m depth from the surface, was found on the sample annealed at 1400 \degree C for 3 h. Further experimental studies are necessary to understand the mechanism behind the extended \qz-rise and the exceptionally high \qz.

Additional studies will also focus on the reproducibility of the techniques to obtain high \qz-values by repeating the 1400 \degree C HT on the same cavity and by applying it to newly built cavities. The role of titanium oxide and hydrogen in this process will be investigated further by applying the 1400 \degree C HT to cavities with Nb flanges. In addition, the process will be applied to fine-grain cavities as well, for comparison. 
Because the breakthrough high-$Q_0$ result we report has such relevance for accelerators application, it is likely that it will trigger new efforts to pursue similar processing techniques and studies within the SRF community.

\begin{acknowledgments}
The authors would like to acknowledge Dr. Peter Kneisel and Dr. Xin Zhao of Jefferson Lab and Dr. John P. Wallace of Casting Analysis Corp. for many valuable discussions. At Jefferson Lab, the authors would like to acknowledge Dr. Andrew Hutton and Dr. Robert Rimmer for supporting this project and Dr. Charles Reece for supporting the XRD measurements at NSU. This manuscript has been authored by Jefferson Science Associates, LLC under U.S. DOE Contract No. DE-AC05-06OR23177. The U.S. Government retains a non-exclusive, paid-up, irrevocable, world-wide license to publish or reproduce this manuscript for U.S. Government purposes.

The electron microscopy was accomplished at the Electron Microscopy Center for Materials Research at Argonne National Laboratory, a U.S. Department of Energy Office of Science Laboratory operated under Contract No. DE-AC02-06CH11357 by UChicago Argonne, LLC.
\end{acknowledgments}

\bibliography{GigiBibliography}

\end{document}